\documentclass[fleqn,usenatbib]{mnras}

\usepackage{newtxtext,newtxmath}

\usepackage[T1]{fontenc}

\DeclareRobustCommand{\VAN}[3]{#2}
\let\VANthebibliography\thebibliography
\def\thebibliography{\DeclareRobustCommand{\VAN}[3]{##3}\VANthebibliography}

\usepackage{graphicx}	
\usepackage{amsmath}	
\usepackage{caption}
\usepackage{subcaption}

\newcommand{\cow}{{AT\,2018cow}}

\title[AT2018cow at late times]{Late time \textit{HST} UV and optical observations of AT~2018cow: extracting a cow from its background}

\author[Anne Inkenhaag et al.]{%
Anne Inkenhaag$^{1,2}$\thanks{E-mail: a.inkenhaag@astro.ru.nl},
Peter~G.~Jonker$^{1,2}$,
Andrew~J.~Levan$^{1,3}$,
Ashley~A.~Chrimes$^{1}$,
Andrew~Mummery$^{4}$,
\newauthor
Daniel A. Perley$^{5}$, 
Nial~R.~Tanvir$^{6}$
\\
$^{1}$Department of Astrophysics/IMAPP, Radboud University Nijmegen, P.O.~Box 9010, 6500 GL Nijmegen, The Netherlands\\ 
$^{2}$SRON, Netherlands Institute for Space Research, Niels Bohrweg 4, 2333~CA, Leiden, The Netherlands\\
$^{3}$Department of Physics, University of Warwick, Gibbet Hill Road, Coventry, CV4 7AL, UK\\
$^{4}$Oxford Astrophysics, Denys Wilkinson Building, Keble Road, Oxford, OX1 3RH, United Kingdom\\
$^{5}$Astrophysics Research Institute, Liverpool John Moores University, IC2, Liverpool Science Park, 146 Brownlow Hill, Liverpool L3 5RF, UK\\
$^{6}$School of Physics and Astronomy, University of Leicester, University Road, Leicester, LE1 7RH, UK
}

\date{Accepted XXX. Received YYY; in original form ZZZ}

\pubyear{2023}

\begin{document}
\label{firstpage}
\pagerange{\pageref{firstpage}--\pageref{lastpage}}
\maketitle

\begin{abstract}
The bright, blue, rapidly evolving AT2018cow is a well-studied peculiar extragalactic transient. Despite an abundance
of multi-wavelength data, there still is no consensus on the nature of the event. We present our analysis of three epochs of
\textit{Hubble Space Telescope (HST)} observations spanning the period from 713-1474 days post burst, paying particular attention
to uncertainties of the transient photometry introduced by the complex background in which AT2018cow resides. Photometric measurements show evident fading in the UV and more subtle but significant fading in the optical. During the last \textit{HST} observation, the transient’s optical/UV colours were still bluer than those of the
substantial population of compact, young, star-forming regions in the host of AT2018cow, suggesting some continued transient
contribution to the light. However, a compact source underlying the transient would substantially modify the resulting spectral
energy distribution, depending on its contribution in the various bands. In particular, in the optical filters, the complex, diffuse background poses a problem for precise photometry. An underlying cluster is expected for a supernova occurring within a young stellar environment or a tidal-disruption event (TDE) within a dense older one. 
While many recent works have focused on the supernova interpretation, we note the substantial similarity in UV light-curve morphology
between AT2018cow and several tidal disruption events around supermassive black holes. Assuming AT2018cow arises from a TDE-like event, we fit the late-time emission with a disc
model and find $M_{BH} = 10^{3.2\pm0.8}$~M$_\odot$. Further observations are necessary to determine the late-time evolution of the
transient and its immediate environment.

\end{abstract}




\section{Introduction}

Multi-wavelength, wide field-of-view surveys at various wavelengths have transformed transient astrophysics. From X-rays with {\em Swift} (\citealt{Burrows2005}) and eROSITA (\citealt{Predehl2021}) through to optical with e.g., the Zwicky Transient Facility (ZTF); \citealt{Bellm2019}, the All-Sky Automated Survey for Supernovae (ASASSN)\footnote{\url{https://www.astronomy.ohio-state.edu/asassn/}}; \citep{Shappee2014}, and the Asteroid Terrestrial-Impact Last Alert System (ATLAS); \citep{Tonry2011} and radio surveys (e.g., the VLA sky survey \citealt{2020PASP..132c5001L}, the Canadian Hydrogen Intensity Mapping Experiment[CHIME]; \citealt{CHIME2022}, and MeerKAT; \citealt{MeerKAT2016}), we can now identify and follow hundreds to thousands of transients, such as gamma-ray bursts, supernovae and fast radio bursts, per year. These high rates result from the combination of areal coverage, depth and cadence of these surveys, and the intrinsic volumetric rate and luminosity function of the transients under consideration. Due to these large, high cadence, sensitive surveys, events that are intrinsically rare, or that are numerous but faint, are also being detected. At the extremes of parameter space, we detect events whose nature stretches plausible progenitor models. These events are thus extremely valuable for study in their own right.

One class of such peculiar transients are fast blue optical transients (FBOTs; e.g., \citealt{Drout2014, Arcavi2016, Whitesides2017, Pursiainen2018, Tampo2020, Ho2021}). A handful of FBOTs have been discovered over the last decade: CSS161010 \citep{Coppejans2020}, AT2018lug/ZTF18abvkwla \citep{Ho2020}, AT2020xnd/ZTF20acigmel \citep{Perley2021}, AT2020mrf \citep{Yao2022}, and the well-known example \cow\ \citep{Prentice2018,Perley2019}. Together, these events form their own class of astrophysical transients, although the FBOT properties are heterogeneous, and the nature of the events is still uncertain. This class of events is characterised by fast rise and decay times, high peak luminosities (absolute peak magnitude $ \lesssim -19$), and early spectra dominated by a blue featureless continuum. Multiple models were suggested, such as peculiar supernovae (SNe) and magnetars formed in double neutron star mergers \citep{Drout2014}. In SNe the timescale of Ni$^{56}$ radioactive decay and the diffusion time scale are critical parameters in the light-curve evolution \citep{Arnett1982}. However, these two time scales are too long to explain the rapid decay and high peak luminosity observed for FBOTs \citep{Drout2014, Pursiainen2018}.

\cow\ was the first FBOT discovered in real-time instead of archival searches. The transient rose to peak rapidly ($>$5 mags in $\sim3.5$ days), was extremely bright (L$_{\rm peak} \approx 10^{44}$~erg~s$^{-1}$; \citealt{Prentice2018, Perley2019}) and was detected across the electromagnetic (EM) spectrum.
The host galaxy CGCG137$-$068 has a luminosity distance of 63.0$\pm$4.4~Mpc (redshift z=0.01404$\pm$0.00002) \textbf(SDSS DR6; \citealt{SDSS2008}. The combination of high (peak) luminosity and relativey low distance meant that many telescopes and satellites could observe and detect it, and led to an extensive observational campaign.

Observations of \cow\ showed that the luminosity decay was too fast to be powered by Ni$^{56}$ decay \citep{Margutti2019}. In addition, the photospheric radius stayed hot and small for hundreds of days \citep{Perley2019, Sun2022}. The optical spectra were featureless the first $\sim$20 days; after that period, emission lines of hydrogen and helium appeared \citep{Prentice2018, Margutti2019, Perley2019}. The spectral evolution has some resemblence to the spectral development of SNe~Ibn and IIn \citep{Fox&Smith2019, Xiang2021} although the lines in \cow\,appeared later than usual for those supernovae.  The X-ray luminosity was high (e.g., \citealt{Margutti2019, Kuin2019}) and showed suggestive evidence for the presence of one or more quasi-periodic oscillations (QPOs) \citep{Zhang2022, Pasham2021}. QPOs are regularly seen in accreting systems, and the combination of a high luminosity and the detection of a QPO, if real, would thus suggest \cow\ is caused by an accreting compact object. 

The host galaxy of \cow\ appears to be a face on spiral system, and there are several (at least two) star-forming regions that lie close to (within $\sim 170$~parsec) the (projected) position of \cow. Assuming \cow\ lies in the plane of the host galaxy and not above or below it, this provides suggestive evidence for a link between massive star evolutionary processes and \cow\ \citep{Lyman2020, MM2019}. On the other hand, \cite{Sun2023} suggest that the low extinction in the transient implies that it is more likely on the near side of the disc and is not necessarily embedded in the star-forming regions. It would argue against a link with a massive star progenitor if this is correct.

Combining all the observed properties, the emission of \cow\ most likely comes from an engine-driven explosion \citep[e.g.,][]{Margutti2019, Perley2019}. Multiple models have been proposed for \cow\ (and FBOTs in general), including magnetars \citep[][]{Prentice2018,Mohan2020, Liu2022}, interactions with the circumstellar material \citep{Sandoval2018,Pellegrino2022} and a pre-existing stellar mass BH disrupting or accreting a companion \citep{Metzger2022}. Among the proposed models, the following two are considered most promising:
An engine-powered core-collapse event, where a compact object is formed that accretes progenitor material \citep[][]{Prentice2018,Perley2019,Margutti2019,Mohan2020}, or a tidal disruption event (TDE) of a white dwarf (WD) or main sequence (MS) star by an intermediate mass black hole \citep[IMBH,][]{Kuin2019, Perley2019}. This class of TDEs may naturally explain the fainter and faster evolution compared to classical TDEs (of stars by a supermassive black hole [SMBH]), as well as provide an explanation for the non-nuclear location of the transient \citep{Maguire2020}. However, the IMBH must reside in a dense stellar environment such that two-body relaxation is efficient enough to scatter a white dwarf (or MS star) into the tidal radius within a Hubble time. Such a dense stellar environment is then a requirement for the TDE interpretation to be viable, although previous research does not provide evidence for such an environment  \citep[e.g.][]{Margutti2019}. However, long-lived, luminous emission from \cow\, makes detecting any putative (underlying) stellar cluster difficult.

The { \it Hubble Space Telescope (HST)} observed \cow\ several times over the four-year period since its first detection. Surprisingly, \cite{Sun2022, Sun2023} detected UV-radiation even more than 4 years after the first detection of \cow. This emission is consistent with a hot and bright source and \cite{Sun2022} suggest a massive star progenitor is most likely involved.

In this work, we present our analysis of the late-time \textit{HST} data of \cow\ , spanning three epochs between 713 and 1474~days after the initial detection. The filters range from F225W in the UV to F814W in the red part of the optical. We perform photometry in multiple ways and investigate the influence of the background measurement on the photometry outcome. We also investigate whether the detected emission is from \cow\ itself or the environment and if there are implications from this distinction for the progenitor scenarios. We investigate if the UV properties can be explained under a TDE scenario and what the implications would be.  

All magnitudes are presented in the AB magnitude system unless specified otherwise. Throughout the paper we use H$_0 = 67.8$\,km\,s$^{-1}$\,Mpc$^{-1}$, $\Omega_{\rm m}=0.308$ and $\Omega_{\rm \Lambda} = 0.692$ \citep{Planck2016}.

\section{Data Analysis}

For this work we use observations of AT2018cow by {\em HST} using the Ultraviolet-Visible (UVIS) channel of the Wide Field Camera 3 (WFC3) at three different late-time epochs. The data we use were taken under program IDs 15974, 16179 and 16925 with PIs A.~Levan, A.~Filippenko and Y.~Chen, respectively. The observations are taken 713~days, 1135~days and 1474~days after the first detection of the transient, which we take to be T$_{\rm 0} = 58285.44$ \citep{Perley2019}. We obtain the individual on-the-fly processed images from the Mikulski Archive for Space Telescopes\footnote{\url{https://archive.stsci.edu/}}, these have had flat field and bias corrections applied and have also been corrected for the impact of charge transfer efficiency on the ageing WFC3 CCDs. 

\subsection{Alignment}\label{sec:alignment}
First we combine the individual images using {\sc astrodrizzle} from the python package {\sc drizzlepack} \citep{drizzlepack}\footnote{\url{https://drizzlepac.readthedocs.io/en/latest/astrodrizzle.html}}. Here, we set the final pixel scale to \texttt{final\_scale=0.025} to utilize sub-pixel dithering to obtain more finely sampled images and to better sample the {\em HST} point spread function (PSF). We use default settings for parameters unless mentioned otherwise. 
Next, we use the {\sc geomap} task in {\sc iraf} \citep{Tody1986, Tody1993} to align the images obtained in the four different filters 713~days after the onset. The sources used for this alignment are the galaxy nucleus \{R.A.,Dec\}=\{16:16:00.582,+22:16:08.286\} and a star \{R.A.,Dec\}=\{16:15:59.147,+22:15:58.88\}: both are detected in all four filters. After this, we use {\sc xregister} to align each filter image obtained at the one (F225W and F336W) or two (F555W and F814W) other epoch(s) to their respective image obtained 713~days after the transient's first detection. We cannot use {\sc xregister} to align all images across all filters because it uses cross correlation to calculate a shift, which does not work well if there are many sources that are not detected in both images, which is the case here when using observations obtained in different filters. The alignment shifts from {\sc geomap} and {\sc xregister} are used to redrizzle the images with an additional shift so the sources align pixel wise in the final images.

\subsection{Aperture Photometry}\label{sec:apphot}

We perform aperture photometry using a circle with a radius of 0.08~arcsec on all the images using dual-image mode in {\sc source extractor} \citep{Bertin1996}, except our detection image, F336W at T=713~days, for which we use single image mode. In dual-image mode source detection is done on one image and the measurements are done on the second image. This enforces the use of a fixed position of the aperture across the different filter images. Using dual-image mode prevent us from having to cross match the detected sources between images and forces {\sc source extractor} to perform photometry at the position of \cow. The choice of aperture radius (corresponding to a diameter of $\sim2$ times the Full Width at Half Maximum (FWHM) ensures we measure most of the emission from \cow\ without measuring too much background. 

We use the drizzled F336W image at epoch 713~days as our source detection image, because there clearly is still emission at the transient location, and more sources are detected in the F336W than in the F225W image. For the photometry we use default values as mentioned in the {\sc source extractor} manual\footnote{\url{https://sextractor.readthedocs.io/en/latest/index.html}} for parameters not mentioned here and adjust parameters such as the FWHM and pixel scale (0.08~arcsec and 0.025~arcsec/pixel, respectively). We set the detection and analysis thresholds to 3.0 sigma to balance between minimizing contamination from spurious detections of hot pixels and allowing the detection of faint sources in the final output. We subtract the local background from the transient light in the final photometry.

Since the individual images are shifted with respect to each other because of drizzling, certain features such as bad pixels or pixels with cosmic rays removed can influence the quality of the signal in multiple pixels in the final combined image (i.e., the noise in the final pixels is correlated to some degree). This can influence the final photometry, which we take into account by using a weight map (\texttt{WEIGHT\_TYPE = MAP\_WEIGHT}) in {\sc source extractor}. This weight map tells {\sc source extractor} which redrizzled pixels contain bad pixels from the individual images, which improves source detection and error estimation, see the {\sc source extractor} user manual for full details. We use the weight map that is produced by {\sc astrodrizzle} during the combination process. 

Aperture corrections are done using appropriate values from the table provided on the WFC3 handbook website\footnote{\url{https://www.stsci.edu/hst/instrumentation/wfc3/data-analysis/photometric-calibration/uvis-encircled-energy}} using r=0.08~arcsec values in the UVIS2 table. For comparison to \cite{Sun2022} we report Vega magnitudes based on the zeropoints from the WFC3 instrument handbook\footnote{\url{https://www.stsci.edu/hst/instrumentation/wfc3/data-analysis/photometric-calibration/uvis-photometric-calibration}}. Photometry is corrected for Galactic foreground reddening following \citet{S&F2011}.

\subsection{PSF photometry}\label{sec:psf_phot}

We also perform PSF photometry to examine whether the source is point-like or extended. We start by cutting out an image (17 by 17 pixels) away from the host galaxy containing an isolated point source (centred on \{RA, Dec\}=\{16:15:59.254, +22:1621.733\} for F555W and F814W, and \{RA, Dec\}= \{16:15:59.148,+22:15:58.379\} for F225W and F335W). This point source is used to provide an estimate of the PSF. Although it does not have as high a signal-to-noise ratio as the computed PSFs available, the advantage of this approach is that it measures the PSF directly on the image. Since the star is much brighter than the transient, the impact of photometric noise on the template PSF is minimal. 

We now proceed to measure the magnitude of a point source at the location of AT2018cow within our images. We linearly interpolate the template PSF to enable sub-pixel centroiding, confirm this model subtracts out cleanly from the PSF star image, and then perform a fit using the pixels with a central position $<6.1$~pixels from the best fitting (x,y) position determined before. This best-fit position of \cow\ is obtained using a 4-parameter fit on the F225 image at T=1474~d (the highest signal-to-noise value of the four UV images), in which the (x,y) position, the PSF normalisation, and the background are left free to vary. The best-fit (x,y) coordinates are then used as fixed input parameters for the fits on the other images (which is possible because of the pixel-wise alignment described in Section~\ref{sec:alignment}), leaving a 2-parameter fit (the normalisation and background are the remaining free parameters). We minimize the $\chi^2$ in this area and report the values for the best-fit background and PSF normalisation.

To produce PSF subtracted images, the PSF template multiplied by the best-fit normalisation is subtracted from the data at the best-fit position. To calculate the magnitude of the subtracted point source, we sum the number of electrons/s in the template PSF in a circular area with a 6-pixel radius around the peak of the PSF, and multiply by the best-fit normalisation. We determine the error on the best fitting peak height by performing a two parameter $\chi^2$ fit, leaving the centroid position fixed on the best-fit position allowing only the PSF normalisation and the background to vary. The error on the height is determined using $\Delta\chi^2 = 2.30$. We calculate the error on the magnitude by multiplying the summed PSF model with the error on the PSF normalisation.

We also perform PSF photometry using {\sc dolphot}(v2.0, \citealt{Dolphin2000}). This software package is specifically designed for PSF photometry in crowded fields. It performs photometry on the individual {\sc \_flc} images and combines the individual measurements into a final (Vega) magnitude for each source. We transform the Vega magnitudes into AB magnitudes using the same difference in zeropoints as mentioned in Section~\ref{sec:apphot}. We use {\sc tweakreg} from {\sc drizzlepac} to align all {\sc \_flc} images to the drizzled F336W T=713~days image, as this has the sharpest PSF. We then perform PSF photometry for this epoch leaving {\sc dolphot} free to search for sources and use the output positions of this run as fixed positions for the other filters and epochs using the "warmstart" option in {\sc dolphot}.

\subsection{Aperture photometry on difference images}\label{sec:im_sub}

We compute difference images using {\sc hotpants} (v5.1.11; \citealt{Becker2015}) by subtracting epoch 3 from epoch 1 or 2 to investigate the brightness of any residual emission at the position of \cow. To perform the subtraction we use default values for the input parameters of {\sc hotpants} except for bgo, ko, and the nsx and nsy parameters where we use values of 0.1, 0.05, 5 and 5, respectively. The parameters (bgo, ko, nsx, nsy) are the spatial orders of the background and kernel variations and the number of stamps within a region in x and y direction, respectively. We also change the gain (which is equal to the exposure time for the \textit{HST} reduced data), and values for the upper and the lower valid data counts for each combination of images we compute a difference image for. We maximize the size of the difference image which is however limited by the need to avoid gaps between the CCDs in the different exposures.
We also perform aperture photometry on these difference images in all filters, using the procedure described below. 

We measure the flux density of any residual on the difference images by determining the number of electrons/s in a circular aperture of 0.08~arcsec radius centered on the position of \cow. From this, we subtract the mean background and we convert to magnitudes. To determine the mean and standard deviation of the background flux density in the difference images, we randomly place circular apertures of the same radius as above within 30 pixels of the position of \cow. In placing these apertures we avoid regions where in the images bright objects are present (see Figure~\ref{fig:background} for an example of the placement of these regions in the epoch 1 F555W image). We find a large spread in the value of the background (on average a factor $\sim1.5$ for the optical filters and between a factor $\sim2$ and $\sim33$ for the UV filters), and therefore the magnitude and its uncertainty depend on the flux density in the background. We will come back to this in the Discussion, while in the paper we use the median background to determine the source magnitude in the difference image and the standard deviation on the background as the $1\sigma$ uncertainty on the magnitude in the difference image. 

If the measured number of electrons/s in the aperture at the position of \cow\ is lower than the mean background, or of similar value to the standard deviation of the background we determine a $3\sigma$ upper limit. For this, we measure the number of electrons/s in a circular aperture with 0.08~arcsec radius centered on the position of \cow\, and we added three times the standard deviation on the background as described above. 
The signal-to-noise ratio of the detection of a source in the difference images is determined as the flux density in the source divided by the standard deviation in the flux density in the background.

\section{Results}

\subsection{Astrometry}\label{sec:astrometryresults}
We find a frame-to-frame alignment uncertainty of $0.005-0.024$~arcsec ($0.19-0.97$~pixels), depending on which combination of frames is looked at. The alignment between images using the same filter is systematically better than alignment between images using different filters. 

A relevant question relating to the nature of the late time emission is whether it is dominated by a point-like component that may be due to the transient, or whether it could arise from an underlying compact region. We therefore check if the position of any emission detected in the difference images is consistent with the position of \cow.

To investigate this we map the early time UV observations (in particular the F336W data) to the later time F555W observations using 10 compact sources which are likely star forming regions within the host galaxy (see Table~\ref{apx:astrometry} for the positions of these sources). We then fit a geometric transformation using {\sc geomap}, allowing only for a shift in position. The centroid locations for the UV source at 713 days and the compact source in F555W at 1474 days are entirely consistent  
($\delta(x) = 0.19 \pm 0.25$ pixels and $\delta(y) = 0.01 \pm 0.19$ pixels). Furthermore, the location of a faint residual visible in the F555W difference image between epoch 1 and epoch 3 is also consistent with the brightest pixel in all epochs of F555W imaging ($\delta(x) = 0.30 \pm 0.36$ pixels and $\delta(y) = 0.06 \pm 0.36$ pixels, where the additional uncertainty arises from the centroid error of the residual emission in the F555W image).

\subsection{Photometry}

\begin{table*}
\caption{The result of our aperture and difference image photometry for AT2018cow, using a circular aperture of r=0.08~arcsec radius as well as PSF photometry following either our manual procedure (see Section~\ref{sec:psf_phot} for details) or using {\sc dolphot}. "Diff.~image" refers to the image obtained after subtracting the image obtained during the third epoch from the epoch 1 or 2 image under consideration (see Section~\ref{sec:im_sub} for details).  Aperture photometry is performed on the diff.~images. Values include aperture correction and a Galactic reddening correction as mentioned in the text. To correct for Galactic extinction we have used A$_{\rm F225W} = 0.524$, A$_{\rm F336W} = 0.334$, A$_{\rm F555W} = 0.214$ and A$_{\rm F814W} = 0.115$. Values without error bars are $3\sigma$ upper limits.  }
\label{tab:mags}
\hspace*{-1.0cm}\begin{tabular}{ccccccccccccc}
\hline
 Filter & Epoch  & \# of &Exp.~time & Aperture phot.&  Aperture phot. & Diff.~image $^\dagger$ & Diff.~image $^\dagger$ & PSF phot. & PSF phot. & {\sc dolphot} & {\sc dolphot}\\
 & (day) & exp. & (sec) & F$_\nu$ ($\mu$Jy) &  (mag) & F$_\nu$ ($\mu$Jy)&  (mag) & F$_\nu$ ($\mu$Jy) & (mag) & F$_\nu$ ($\mu$Jy) & (mag) \\
\hline 
F225W & 713 & 3 & 1116 & $1.17\pm0.06$ & 23.73$\pm$0.05 & $0.45\pm0.06$ & $24.77\pm0.14$ & $1.41\pm0.11$ & $23.53\pm0.09$ & $1.08\pm0.06$ & $23.82\pm0.06$\\
F336W & 713 & 3 & 1116 & $0.87\pm0.04$ & 24.05$\pm$0.04 & $0.28\pm0.04$& $25.28\pm0.15$ & $0.82\pm0.07$ & $24.11\pm0.09$ & $0.75\pm0.03$ & $24.21\pm0.05$\\
F555W & 713 & 3 & 1044 & $0.39\pm0.02$ & 24.92$\pm$0.04 & $0.09\pm0.02$& $26.54\pm0.25$ & $0.48\pm0.06$ & $24.69\pm0.13$ & $0.27\pm0.06$& $25.32\pm0.06$\\
F814W & 713 & 3 & 1044 & $0.37\pm0.03$ & 24.97$\pm$0.09 & $0.11\pm0.04$& $26.3^{+0.4}_{-0.3}$ & $0.14\pm0.06$ & $26.0^{+0.6}_{-0.4}$ & $0.13\pm0.2$ & $26.06\pm0.17$\\
\hline
F555W & 1135 & 2 & 710 & $0.33 \pm 0.02$ & 25.10$\pm$0.06 & $<0.09$& $>26.5$ & $0.35\pm0.07$ & $25.07\pm0.22$ &$0.18\pm0.02$ & $24.79\pm0.10$\\
F814W & 1135 & 2 & 780 & $0.23 \pm 0.03$ & 25.48$\pm$0.15 & $<0.11$& $>26.3$ & $0.10\pm0.06$ & $26.4^{+1.1}_{-0.5}$ & $0.05\pm0.02$ & $27.2^{+0.7}_{-0.4}$\\
\hline
F225W & 1474 & 3 & 1845 & $0.71\pm0.04$ & 24.28$\pm$0.06 & $-$ & $-$ & $0.76\pm0.08$ & $24.20\pm0.11$ & $0.54\pm0.04$ & $24.57\pm0.07$\\
F336W & 1474 & 3 & 1953 & $0.61\pm0.02$ & 24.44$\pm$0.04 & $-$ & $-$ & $0.54\pm0.04$ & $24.56\pm0.08$ & $0.51\pm0.02$ & $24.63\pm0.04$\\
F555W & 1474 & 3 & 1149 & $0.32\pm0.01$ & 25.15$\pm$0.05 & $-$ & $-$ & $0.37\pm0.05$ & $24.98\pm0.15$ & $0.19\pm0.01$ & $25.68\pm0.07$\\
F814W & 1474 & 3 & 2271 & $0.29\pm0.02$ & 25.24$\pm$0.07 & $-$ & $-$ & $0.08\pm0.04$ & $26.6^{+0.6}_{-0.4}$ & $0.08\pm0.01$ & $26.53\pm0.17$\\
\hline 
\end{tabular}
\newline $^\dagger$This implicitly assumes that any light at the position of the transient at epoch 3 is not due to \cow.
\end{table*}

\subsubsection{Aperture photometry}
The results of our aperture photometry can be found in Table~\ref{tab:mags}. In the two UV filters (F225W and F336W) and the F555W filter the source has faded between the first and the third epoch (by $0.55\pm0.08$,  $0.39\pm0.06$ and 0.23$\pm$0.06 magnitudes, respectively). In the F814W band the magnitudes are consistent with being the same within 3$\sigma$.

\subsubsection{Photometry from PSF fitting}

\begin{figure*}
 \centering
 \hspace*{-.5cm}\includegraphics[width=0.5\textwidth]{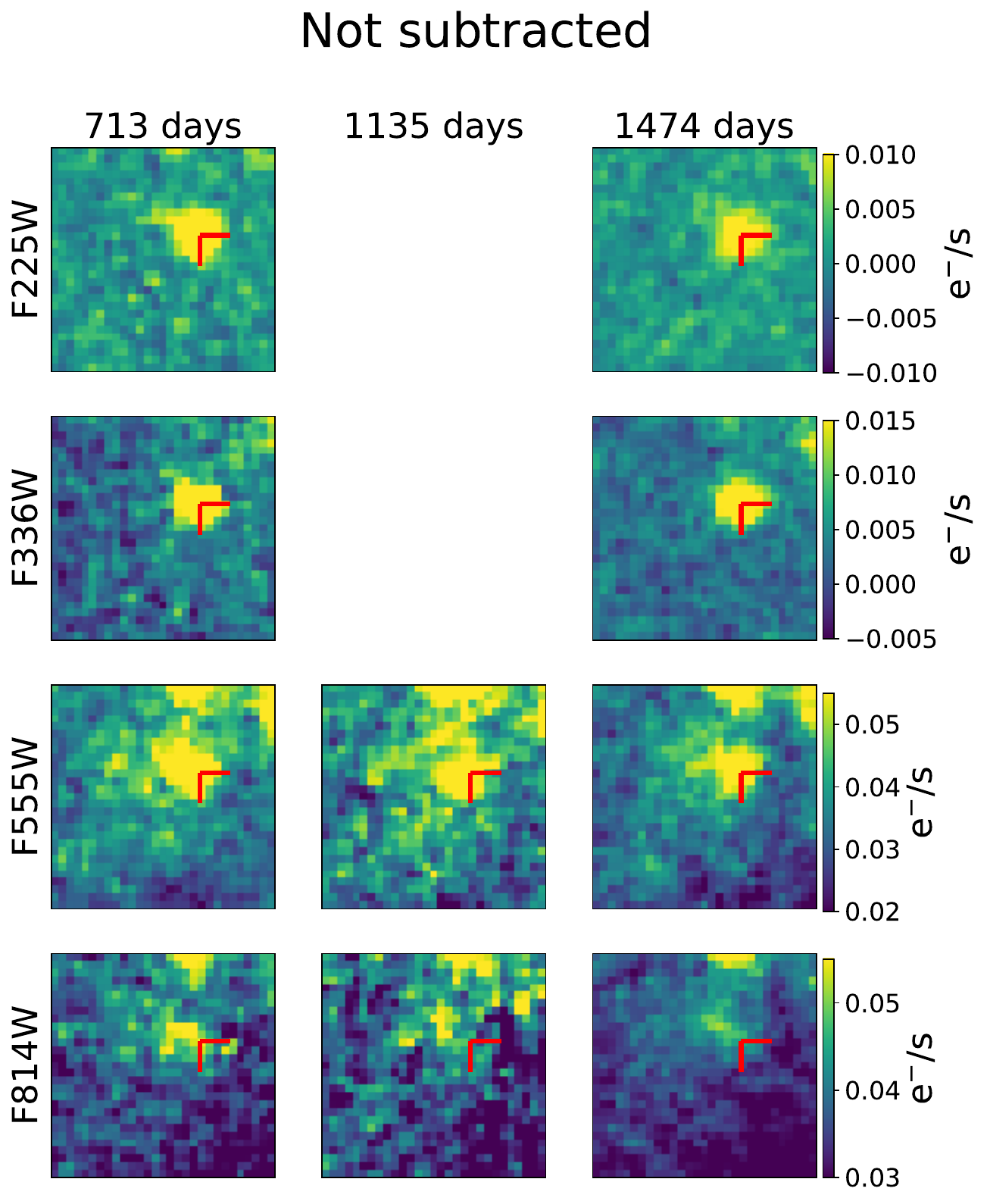}
 \hspace*{+.2cm}\includegraphics[width=0.5\textwidth]{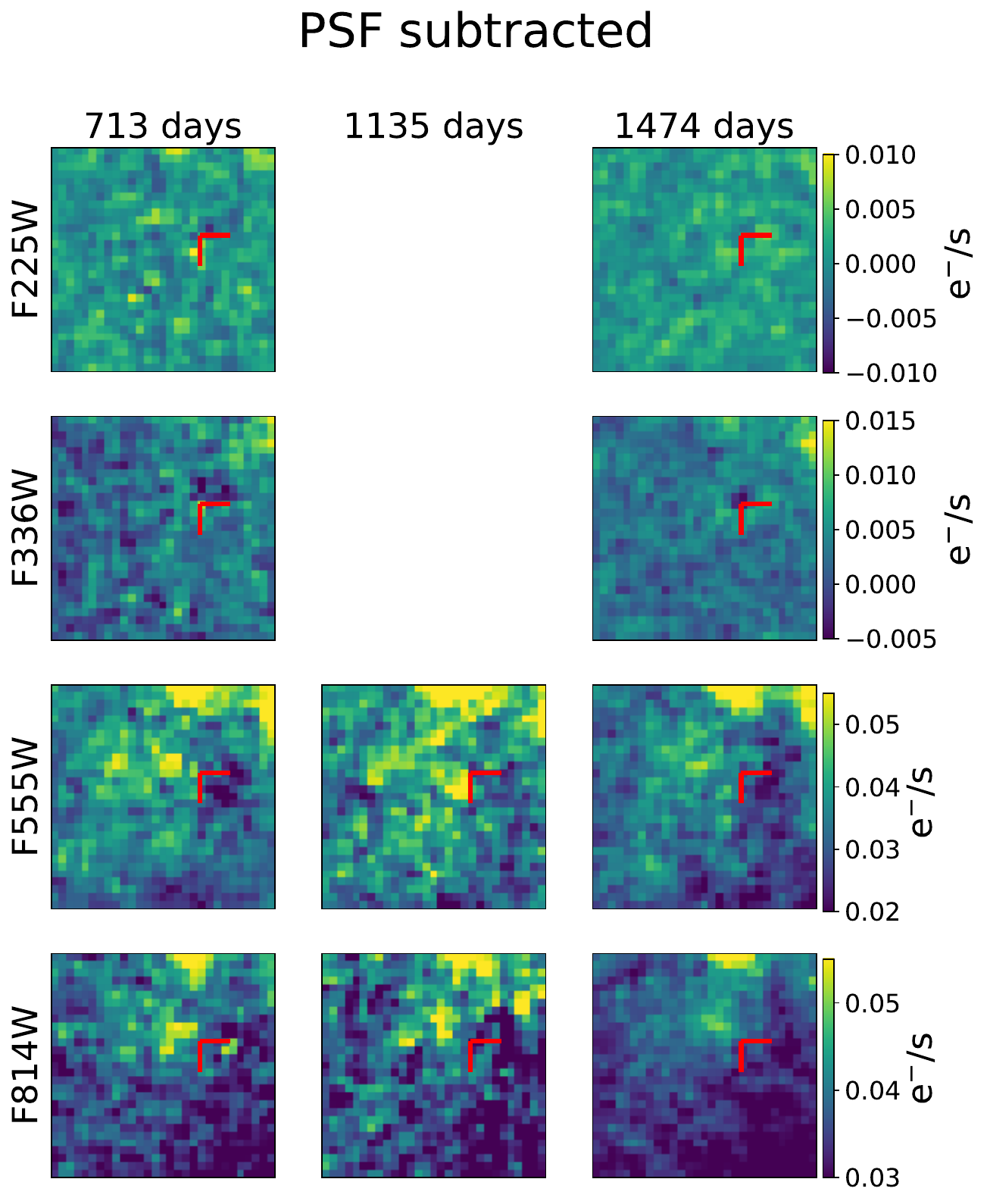}
  \caption{{\it Left panel}: Three columns of four rows of cutout images close to the location of \cow\ for all filters (rows) and epochs (columns). Intensity is given in e$^-$/s in a colour scale, with blue being the least intense and yellow most intense. The best fit centroid position of the PSF to the emission at the location of \cow\ lies where the two red lines touch. The cross hairs have a length of 0.1 arcsec. {\it Right panel}: Three columns of four rows of cutout images showing the residuals of PSF subtraction at the location of \cow\ for all filters (rows) and epochs (columns). The exposure times for the last epoch is longer than for the first two epochs, the second epoch having the shortest exposure time of all, which explains the difference in noise properties in the residual images.}
 \label{fig:PSFresiduals}
\end{figure*}

In the {\it right panels} of Figure~\ref{fig:PSFresiduals} we show the residuals after PSF subtraction in high contrast for all epochs and filters. The best-fit position of the centroid of the PSF model (as determined on the F225W T=1474~days image) is marked by red pointers in each image. The {\it left panels} show the same images, before subtracting the best-fit PSF model. In general, the emission in the UV filters subtracts out well while the point source subtraction in the optical filters reveals/highlights the presence of residual emission close to and in some cases under the source position. The magnitudes of the subtracted point sources are listed in Table~\ref{tab:mags} under PSF photometry. We find reduced $\chi^2$ values between 0.5 and 1.1 for the best fits of the PSF normalisation and background value, showing our model describes the data well. Generally, the PSF magnitudes of the subtracted point source are consistent within $3\sigma$ with those derived through our aperture photometry for all filters, although the PSF magnitudes in the F814W filter are systematically fainter (but still consistent within 3$\sigma$). 

Any small residuals present in the PSF-subtracted images obtained through the UV filters can be explained by the fact that the PSF in the UV varies as a function of source location on the image. Due to various factors (such as the coatings of the optical elements) the UV PSF contains broader wings than the optical PSF and these broad wings have complex features\footnote{\url{https://hst-docs.stsci.edu/wfc3ihb/chapter-6-uvis-imaging-with-wfc3/6-6-uvis-optical-performance}}. Since we try to fit the central part of the PSF to the data, the features in the wings can leave residuals when a template PSF determined at one place of the image is subtracted from a source at another location in the image.

\subsubsection{Photometry using {\sc dolphot}}

The results or our PSF photometry using {\sc dolphot} can be found in Table~\ref{tab:mags}. However, {\sc dolphot} yields no detection at the position of \cow\ in F814W for any of the observation epochs and in F555W at the epoch at T=1135~days, unless the input source position is fixed, as described in Section~\ref{sec:psf_phot}, which is effectively equivalent to forced photometry at the position of \cow.

\subsubsection{Aperture photometry on the difference images} \label{sec:diff_images}

\begin{figure*}
 \centering
 \hspace*{-1.cm}\begin{subfigure}[b]{0.52\textwidth}
    \centering
    \includegraphics[width=\textwidth]{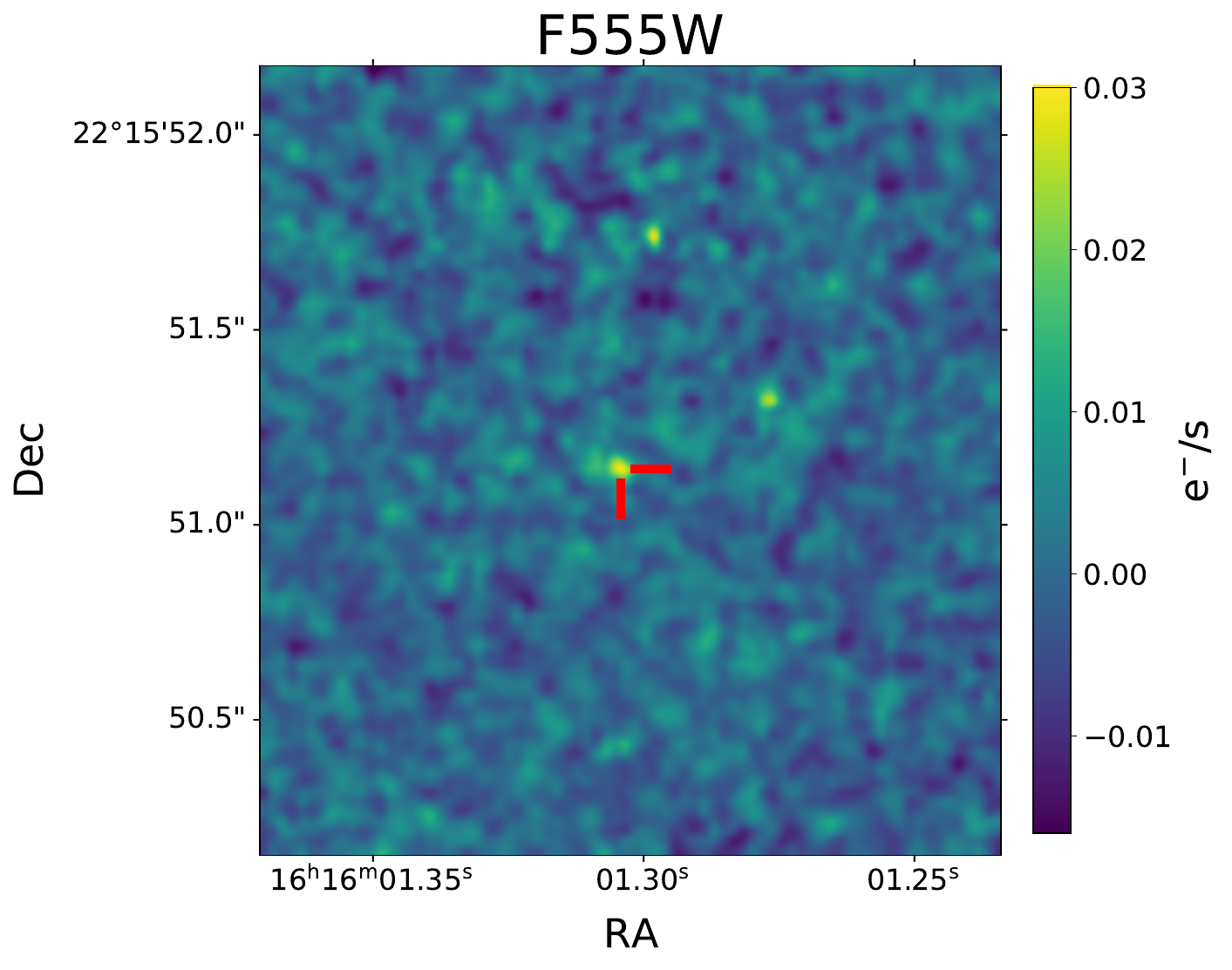}
 \end{subfigure}
 \hfill
 \hspace*{-.5cm}\begin{subfigure}[b]{.53\textwidth}
    \centering
    \includegraphics[width=\textwidth]{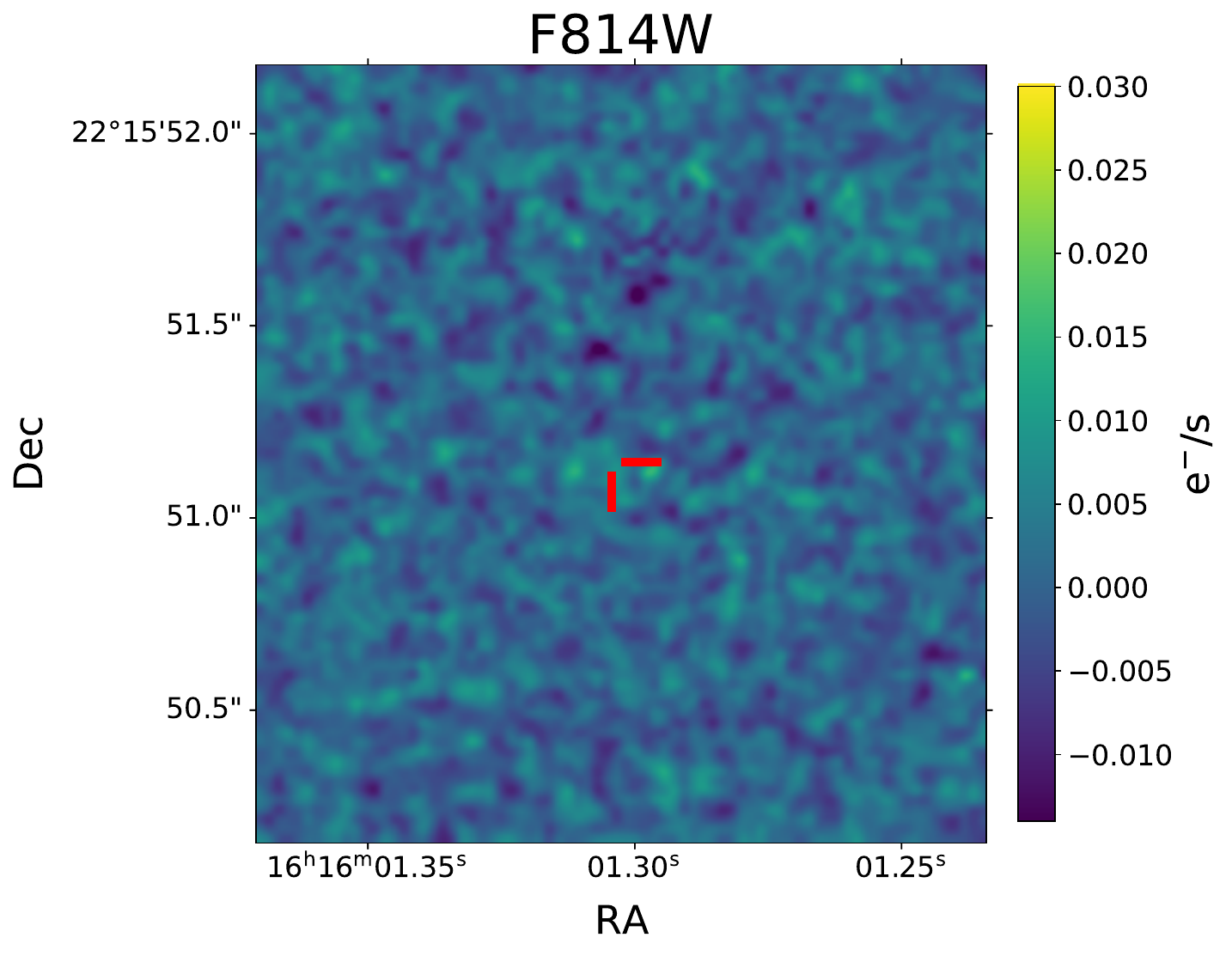}
 \end{subfigure}
    \caption{The residual flux after subtracting the image obtained at T=1474~d from the T=713~d image for the two optical filters (F555W {\it left panel}; F814W {\it right panel}) using {\sc hotpants}. The location of \cow\ is indicated with red thick marks. Residual emission is present at the position of \cow\ with a signal-to-noise of $4.5$ in the F555W difference image and signal to noise of $3.4$ in the F814W difference image ({\it left panel}; see Section~\ref{sec:diff_images} of the main text for details.)}
    \label{fig:diff_images}
\end{figure*}

Figure~\ref{fig:diff_images} shows the difference images created by subtracting the epoch 3 images from the epoch 1 images for the two optical filters. Here, the position of \cow\ is indicated by red markers. In the F555W difference image ({\it left panel}) there is residual emission near the position of \cow. This residual emission is not an artefact due to uncertainties in alignment as there are no such residuals at the positions of other point sources in the difference image. This residual is detected at a signal-to-noise ratio of $4.5$ with a magnitude of $26.54\pm0.25$, consistent with the difference between the F555W magnitude in epoch 1 and epoch 3 as measured through aperture photometry. 

For the observations obtained in the F814W filter, no distinguishable residual emission is present (when looking by eye) in the difference image, as can be seen in the {\it right panel} of Figure \ref{fig:diff_images}. Following the same procedure as for F555W above we find a signal-to-noise ratio of $3.4$ with a magnitude of $26.3^{+0.4}_{-0.3}$. Subtracting the epoch 3 image and then measuring the flux/magnitude of the residual measures the decaying component in the \cow\, light. An alternative way of looking at the difference image is that it assumes all emission at epoch 3 (T=1474~days) is due to an underlying source at the position of \cow. 
Under "Diff.~image" in Table~\ref{tab:mags}, we list our results for aperture photometry on all difference images created by subtracting the epoch 3 from the epoch 1 and epoch 2 images. For the F555W and F814W epoch 2 minus epoch 3 difference images the measured flux density in the aperture is consistent with that expected due to variations in the background, hence we report $3\sigma$ upper limits of $>26.5$ in F555W and $>26.3$ in F814W.

\subsection{Lightcurve}

\begin{figure*}
 \centering
 \hspace*{-.7cm}\includegraphics[width=0.7\textwidth]{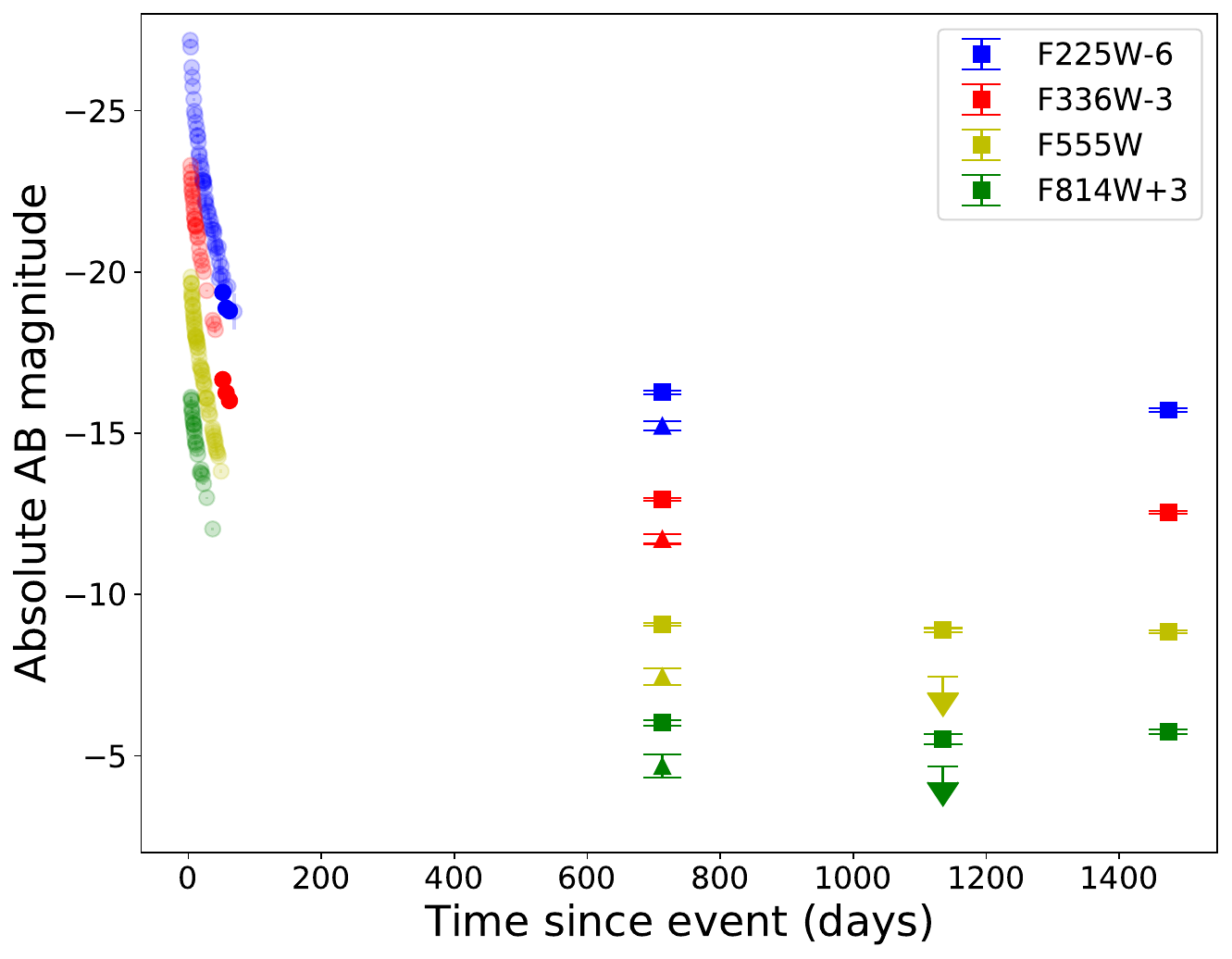}
    \caption{\cow\, light-curves in different filters, F225W in blue, F336W in red, F555W in yellow and F814W in green (with offsets as indicated in the legend). The early time data is from \protect\cite{Perley2019} in transparent circles and \protect\cite{Sun2022} in opaque circles. 
    Our aperture photometry results marked with squares assume all flux measured in the last (third) epoch is due to the transient, whereas for the measurements indicated with triangles and downwards pointing arrows (for upper limits) we assumed that all detected flux in epoch three is unrelated to \cow. The error bars are at a $1\sigma$ confidence level. The horizontal bars through the markers do not indicate uncertainties in the observation date but instead they are the end caps of the error bars on the magnitudes. } 
 \label{fig:lightcurve}
\end{figure*}

Out of the three different ways we used to measure the photometry of \cow\,the aperture and PSF photometry agree within 3~$\sigma$. The aperture photometry on the difference images (epoch 1 or epoch 2 minus epoch 3) yield fainter results for the source brightness.
This can be explained as follows: through photometry on a difference image we are sensitive only to the component of the light that varied between the epochs under consideration. In the extreme scenario that the third epoch contains {\it no light} from \cow\, the magnitudes determined through analysis of the difference images are relevant. In the opposite extreme scenario, we assume that {\it all the light} detected at the third epoch comes from \cow. Clearly, whether either or none of these two is a viable assumption may well depend on the filter of the observations under consideration.

We show the brightness evolution of \cow\ as a function of time in Figure~\ref{fig:lightcurve}, using the results of our aperture photometry on the images and the difference images, together with early time data from \cite{Perley2019} and \cite{Sun2022} (circles). Even though the effective wavelengths of the filters used in the early UVOT and later {\em HST} observations are slightly different, we compare UVOT/UVW1 to {\em  HST}/F235W, UVOT/U to {\em HST}/F336W, UVOT/V to {\em HST}/F555W and UVOT/I to {\em HST}/F814W. Different filters are indicated using different colours and we offset the light-curve in each filter by a constant shown in the figure legend for display purposes. Our aperture photometry measurements are shown with squares and our measurements for \cow\, obtained assuming the third epoch contains no transient emission (aperture photometry on the difference images) are indicated with triangles when a residual was detected or downwards pointing arrows when an upper limit to the source magnitude was determined.
Comparing the early-time ($<100$~days after discovery) measured decay in absolute magnitude with absolute magnitude (limits) obtained for the last three {\it HST} epochs, it is clear that the detected emission is brighter than expected had this trend continued.

\subsection{Comparison of \cow\ and compact UV selected sources}\label{sec:SF_regions}

\begin{figure*}
 \centering
 \hspace*{-.7cm}\includegraphics[width=0.7\textwidth]{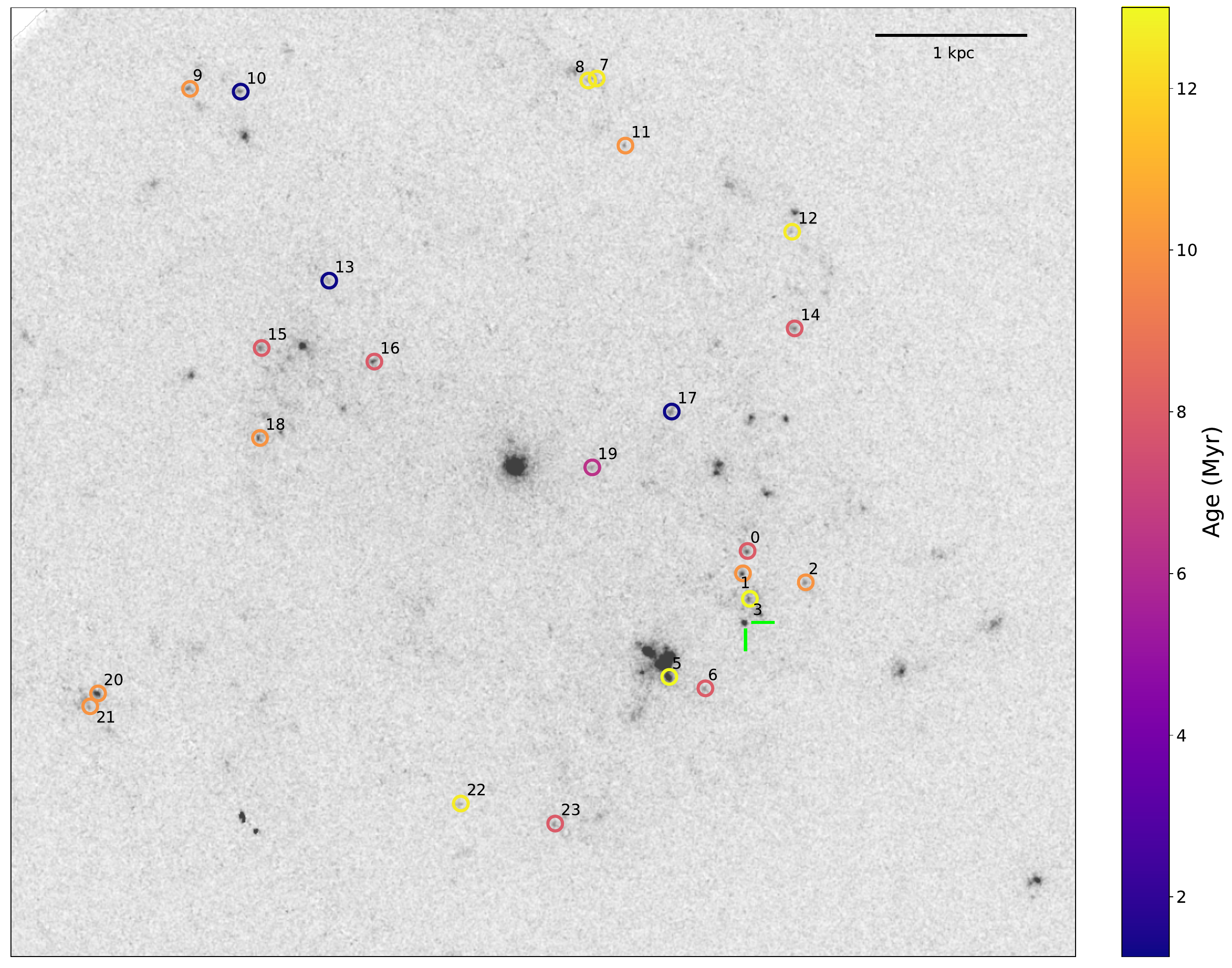}
  \caption{Greyscale image of the host galaxy of \cow\ in the F336W filter at T=713~days., with the ages of compact UV-selected sources that were detected in all four filters indicated by coloured circles. The colours correspond to population ages, indicated by the colour bar and derived from BPASS SED fitting as described in Section \ref{sec:SF_regions}. The location of \cow\ is marked by green cross hairs. Number labels for the regions are as in Table~\ref{tab:bpass}.}
 \label{fig:SF_regions}
\end{figure*}

Next, we explore whether \cow\ is localised in an unusual region of its host galaxy by fitting synthetic spectra of simple stellar populations to 23 compact UV-selected star-forming (SF) regions within the host (plus the location of \cow). These SF regions were selected by running {\sc source extractor} in dual image mode in the same way as for \cow\ (see Section~\ref{sec:apphot}) removing sources that are not detected in all four filters at T=713~days. We also removed sources that are detected with a signal-to-noise ratio $<3$. From these sources we select those that have a constant magnitude (within $3\sigma$) as measured on T=713~days and T=1474~days. Differences in magnitudes between these epochs might be caused by e.g., different orientations of \textit{HST} during the observations. We ignore epoch 2 in this comparison because the exposure time is shorter and there are only two exposures, resulting in a bad removal of cosmic rays. 

Next, we select the sources that behave PSF-like in F336W. We test this by performing aperture photometry using two different values for the radius of the circular aperture and we retained sources only if the difference in their photometry was consistent with the different aperture corrections for a point source given the two different aperture radii. A full list of the positions and magnitudes of the sample can be found in Table~\ref{tab:bpass} in the Appendix~\ref{apx:bpass}.

To determine the ages of these regions, we make use of BPASS v2.2.1 \citep[Binary Population and Spectral Synthesis,][]{Eldridge2017,Stanway2018} synthetic spectra, assuming a single burst of star formation and a metallicity (by mass fraction) of $Z=0.01$ \citep[based on the host metallicity found by][]{Lyman2020}. For each region, SED fitting is performed by convolving the BPASS spectra at each age (52 ages spaced logarithmically from 1\,Myr to 100\,Gyr are used) with the filter response curves for F225W, F336W, F555W, and F814W \citep{2012ivoa.rept.1015R,2020sea..confE.182R}, converting magnitudes to fluxes, and vertically scaling\footnote{The scaling is needed as the synthetic spectra are for a $10^6$~M$_\odot$ population, in Solar luminosity per Angstrom} the synthetic spectra to minimise $\chi^{2}$ across both age and different values for the host-intrinsic extinction. The extinction in each filter is calculated by adopting their effective wavelengths and using the python {\sc extinction} module \citep{barbary_kyle_2016_804967}, with a \citet{Fitzpatrick1999} extinction curve and R$_{\rm V}=3.1$. Galactic extinction is already accounted for as described in Section~\ref{sec:apphot}.

For each region we determine a best-fit age and extinction A$_{\rm V}$. Full results can be found in Appendix~\ref{apx:bpass}. The extinction values are in the range 0.0--0.6 \citep[in broad agreement with A$_{\rm V}=0.2$ as found by][for nearby star forming complexes]{Sun2023}, and the ages range from 6--25\,Myr. These ages are younger than the tens of Myr found by \citet{Lyman2020} for example, but this can be explained by the spaxel size of their MUSE integral field unit data, which averages over larger physical areas than the compact star-forming regions we are probing here. 

The reduced $\chi^{2}$ values (which are the same as the $\chi^2$ values because our fit has one degree of freedom) for the 23 compact star forming regions are typically around $\sim$1--10; whereas the fits at the location of \cow\ (at both 713 and 1474~days) are notably poorer, with $\chi^{2}=47$ and $37$, respectively. The fits at the location of \cow\ favour very little to no extinction, and tellingly, favour the youngest population age available in the BPASS outputs (1\,Myr), whilst still failing to reproduce the extremely blue observed SED. 

In Figure~\ref{fig:SF_regions}, we show the 23 UV-selected star-forming regions over an F555W image of the host galaxy. Each of the 23 regions is encircled, where the colour of the circle corresponds to the best-fit population age. Young stellar populations are present across the galaxy, with no preference for particularly young star forming regions in the vicinity of \cow, although there are 2 star forming regions within $\sim 400$~parsec of the transient (regions 1 and 3, these were unresolved in previous non-{\it HST} data sets).

\subsection{Spectral Energy Distribution of \cow} \label{sec:SED}

\begin{figure*}
 \centering
 \hspace*{-.7cm}\includegraphics[width=0.51\textwidth]{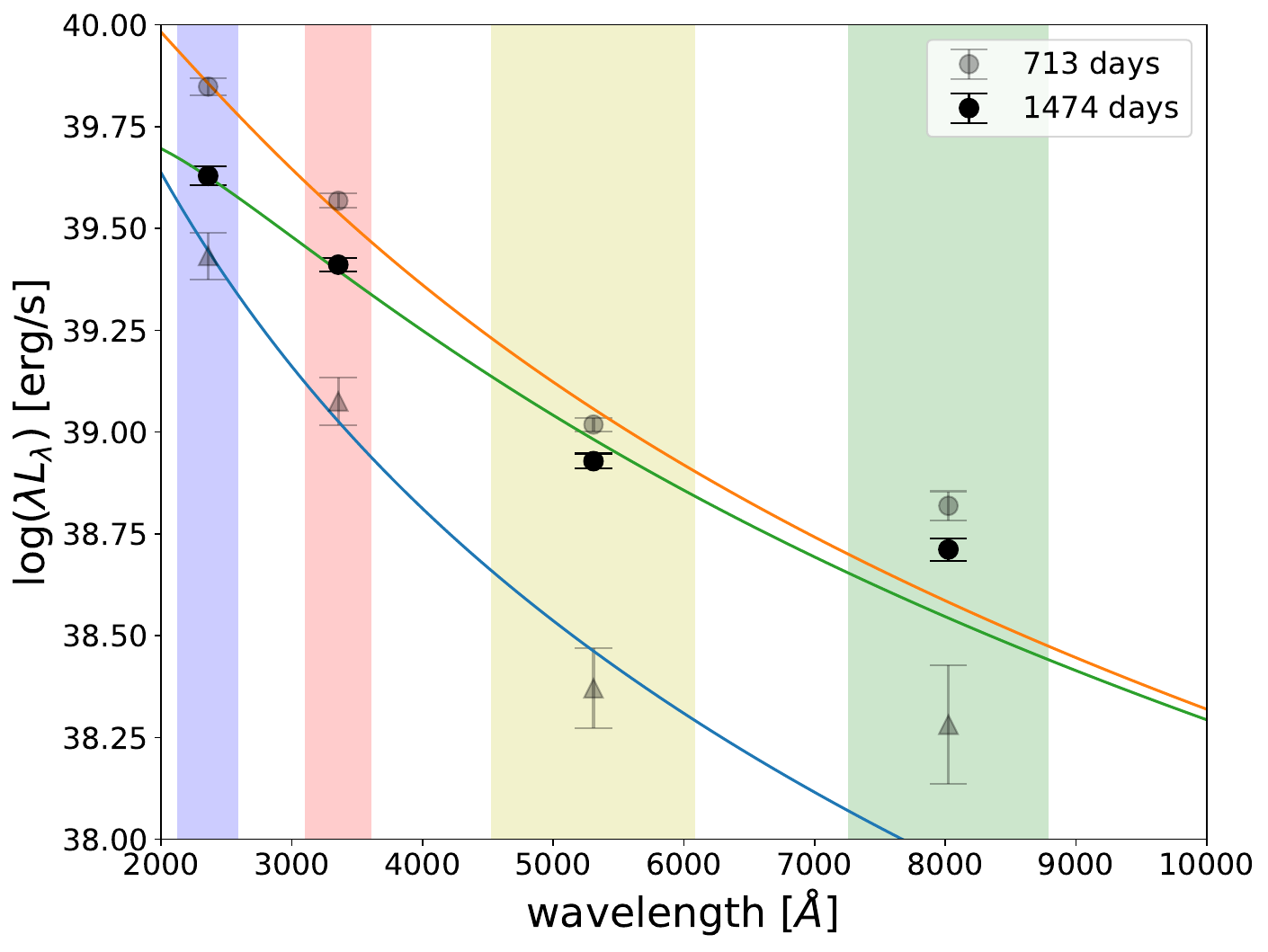}
 \hspace*{-.4cm}\includegraphics[width=0.51\textwidth]{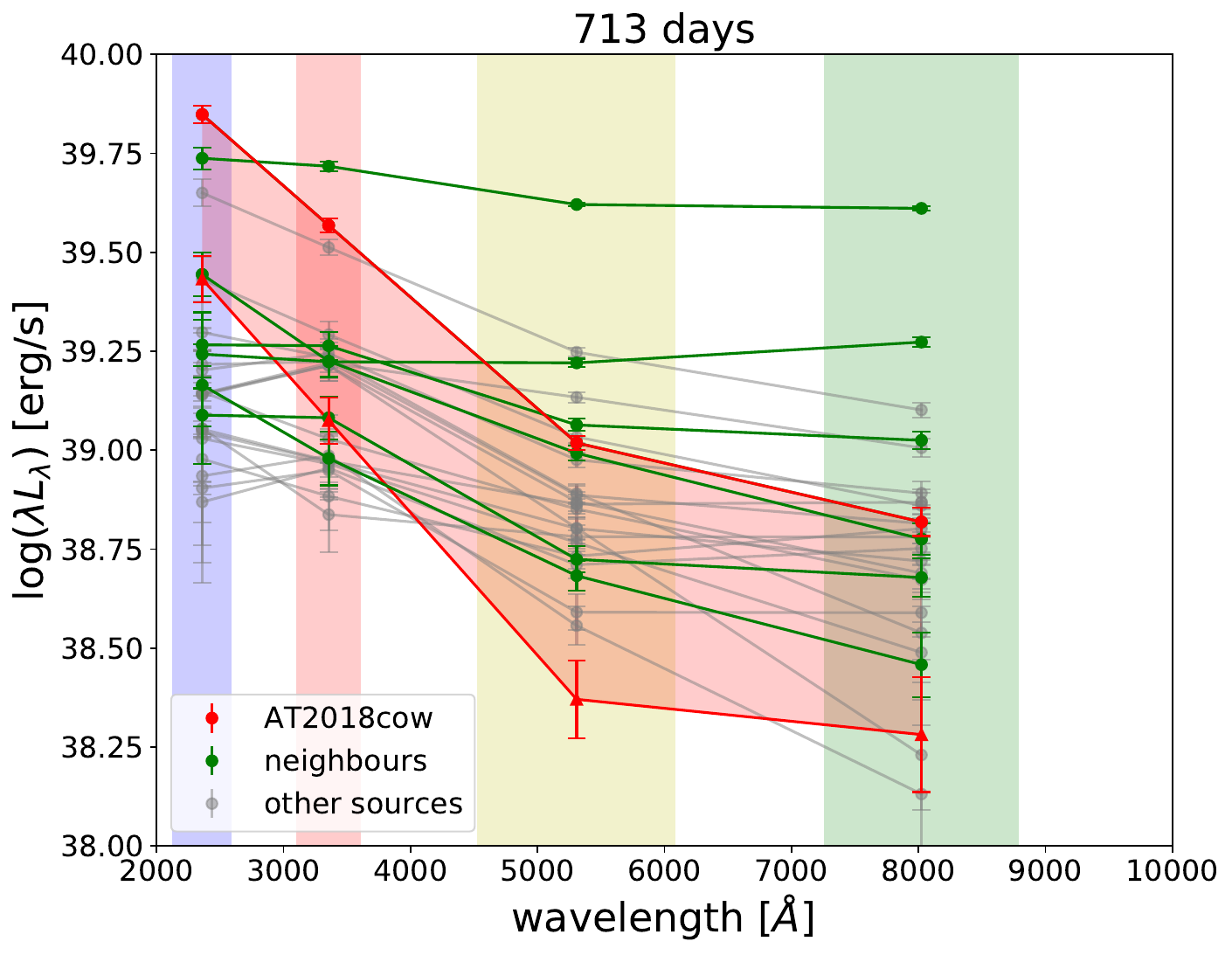}
\caption{{\it Left panel:} The spectral energy distribution (SED) of the emission detected at the position of \cow~at T=713~d and T=1474~d. The four vertical coloured bands are centered on the effective wavelength of the filters used for the observations while the width of the vertical bands indicate the passband rectangular width of the filters. Light grey markers are used for the data obtained at T=713~d. Here, the light grey circles indicate the measured flux density assuming all light in the third epoch (T=1474~d) originates from \cow, whereas light grey triangles are used for measurements obtained assuming none of the third epoch light is due to \cow. The circles are always at a higher flux density than the triangles. The black symbols represent our measurements of the source flux density obtained at T=1474~d. The lines are Planck functions fitted to the four-point SEDs at T=713~d (orange), T=1474~d (green), and to the  grey triangles (blue). The best fitting values for the temperature and the radius, and reduced $\chi^2$ values can be found in Table~\ref{tab:planck}. The fit to the difference image gave unphysical (a negative) values for the temperature when considering the uncertainty on the temperature using both python routines {\sc curve\_fit} and {\sc lmfit}. To obtain an estimate of the uncertainty on the black body temperature we fixed the radius to the best fitting value and determine for which value for the temperature around the best fitting temperature value $\Delta\chi^2=1$. From the reduced $\chi^2$ values and the Figure we conclude that a single black body function is only a reasonably good description of the SED for the light grey triangles. {\it Right panel:} The SEDs of our list of compact UV-detected sources at T=713~d (Table~\ref{tab:bpass} contains selected properties of these sources). The data for \cow~is in red with the marker shapes as mentioned above. We make a distinction between "neighbours" shown in green and "other sources" in light grey. See the main text for the definition of "neighbour" and "other sources". Irrespective of the interpretation of the \cow\ data at T=1474~d, the F555W$-$F225W colour of the source at the position of \cow\, is bluer than any of our compact UV-detected sources.}
\label{fig:SED}
\end{figure*}

\begin{table*}
\centering 
\caption{Results from fitting a Planck black body function to the {\it HST} spectral energy distribution for \cow. These fits are shown in Figure~\ref{fig:SED}.} 
\label{tab:planck}
\begin{tabular}{lccccc}
\hline %
Epoch & log(T (K)) & radius (R$_\odot$) & luminosity (erg s$^{-1}$) & reduced $\chi^2$ & degrees of freedom \\
\hline 
1: T=713~d & $4.54 \pm 0.04$ & $34 \pm 3$ & $(6 \pm 2) \times 10^{39}$ & 17.2 & 2 \\
3: T=1474~d & $4.37 \pm 0.02$ & $43 \pm 2$  & $(1.9 \pm 0.4) \times 10^{39}$ & 17.9 & 2 \\
\hline
Epoch 1 - Epoch 3 & $5.03 \pm 0.04$ & $9 \pm 6$ & ${(4^{+5}_{-4})\times 10^{40\dagger}}$ & 2.2 & 2 \\
\end{tabular}
\newline $^{\dagger}$ See Section~\ref{sec:SED} for the explanation on how the error bars on the luminosity were determined.
\end{table*}

Figure~\ref{fig:SED} ({\it left panel}) shows the spectral energy distributions (SEDs) for \cow\ as measured at epoch 1 (T=713~d) and at epoch 3 (T=1474~d). The black markers represent measurements from the third epoch, while the grey markers those of the first epoch. The marker symbols are the same as in Figure~\ref{fig:lightcurve}. The coloured bands represent the FWHM of the filter with the corresponding colour in Figure~\ref{fig:lightcurve} \footnote{\url{https://hst-docs.stsci.edu/wfc3ihb/chapter-6-uvis-imaging-with-wfc3/6-5-uvis-spectral-elements}}. The {\it right panel} of Figure~\ref{fig:SED} shows both possible extremes of the SED of \cow\ in red compared to the SEDs of compact UV-selected sources detected in a box of 180x180 pixels centred on the position of \cow\ ("neighbours") in green, and "other sources" in the rest of the host galaxy in grey for T=713~d.  From this red shaded region it is clear that for either of the two extreme scenarios for the aperture photometry at epoch T=713~d, the F555W$-$F225W colour of \cow\ is bluer than that of the neighbours. The {\it left panel} of Figure~\ref{fig:SED} shows that the SED for the third epoch lies in between the aperture photometry SED and the difference image SED. Therefore, the T=1474~d SED is also bluer than that of the neighbours.  

We fit a Planck function to the four-filter SEDs at T=713~d, T=1474~d, and to the four-filter SED when assuming none of the third epoch emission is due to \cow, with the best-fit black body over-plotted in the {\it left panel} of Figure~\ref{fig:SED} in orange, green, and blue, respectively. The best-fit values for the temperature and radius, the calculated luminosity, the number of degrees of freedom, and the reduced $\chi^2$ values are presented in Table~\ref{tab:planck}. The error on the temperature for the fit to the epoch 1 - epoch 3 difference image was calculated by fixing the radius to the best-fit value and finding the value for which $\Delta \chi^2=1$. This was done because the error calculated by the fitting algorithm was larger than the best fitting value for the temperature. Only the reduced $\chi^2$ value for the fit to the epoch 1 SED derived assuming epoch 3 contains no light from \cow\ is close to 1 (at a value of 2.2). However, the error on the luminosity is very large due to the large errors on the radius. Due to the sizes of the error bars on the magnitudes obtained with aperture photometry on the difference image, the fit of the Planck function is dominated by the two data points in the UV bands, meaning the fit is almost degenerate for a two-parameter Planck function. This results in a large error on the fit and therefore on the calculated luminosity. 

\section{Discussion}

In this paper, we present aperture and PSF photometry of {\it HST} data of the FBOT AT2018cow. 
We first compare our results in Table~\ref{tab:mags} with the results from the epoch 1--3 PSF photometry by \cite{Sun2022} and \cite{Sun2023}.
We find that our measurements in the UV filters yield a source that is consistent within $3\sigma$ in the first epoch, while in the last epoch our source is brighter than they report (there are no UV images for the second epoch). In the optical filters our measurements indicate a brighter source in all epochs than found by \cite{Sun2022, Sun2023}.  They assumed all the light is emitted by \cow.
Additionally, \cite{Sun2022, Sun2023} find a steeper decay between epoch 1 and 3 in the UV filters ($1.02\pm0.11$~mag and $0.57\pm0.07$~mag for F225W and F336W, respectively) than we do ($0.55\pm0.08$~mag and $0.39\pm0.06$ for F225W and F336W, respectively). Furthermore, they find no evidence for a decay in the source brightness in the optical filters, whereas we do ($0.23\pm0.06$~mags in F555W, and a detection with a signal to noise of 3.4 in the F814W epoch 1 and epoch 3 difference image with a magnitude of $26.3^{+0.4}_{-0.3}$). We will investigate possible reasons for these differences below.

Next, we compare with the epoch 1--3 PSF photometry reported in \cite{Chen2023}.
Our aperture as well as our manual PSF photometry give brighter magnitudes for \cow\, than \cite{Chen2023}, although the difference is small for the two UV filters it increases for the optical filters. Comparing the magnitudes in the \cite{Chen2023} table 1 with their figure 6 we deduced that their table 1 magnitudes are corrected for extinction. However, if they are not, the differences with our extinction-corrected magnitudes is reduced, especially for the UV filters. However, still, only the measurements in F225W (both epochs) and F555W T=1135~days would be consistent withing within the 3$\sigma$ error. Our {\sc dolphot} PSF photometry results are consistent within $3\sigma$ with the values presented by \citet{Chen2023} in their table~1 if those values are not corrected for Galactic extinction. 
When leaving the position as a free parameter, {\sc dolphot} does not find a source in F814W at any epoch and also not in F555W at the epoch T=1135~days. Only forced photometry (i.e. keeping the source position fixed) yields a sometimes marginal detection of the source at those epochs and filters.

However, this does not necessarily mean the photometry presented by \cite{Sun2022, Sun2023}, \cite{Chen2023} or our photometry results are wrong. 
In practice, contributions from other sources besides a point source may influence the measurements, or if no point source is present but if the observed light is dominated by diffuse emission (on the scale of $\sim$few times the PSF size) in \cow's host galaxy galactic disc, PSF photometry provides an upper limit on the magnitude of a point source at the location of \cow. Instead, aperture photometry may over-estimate the true flux density of the transient if the light from the point source and diffuse emission in the galactic disc are of similar magnitude. In practise, the estimated value of the background flux density under \cow\, may influence the determined magnitudes especially in crowded regions like that of \cow. Next, we investigate the potential influence of the choice of the background region used to estimate the flux density at the position of \cow.

\begin{table*}
\caption{The result of our aperture photometry for AT2018cow, using a circular aperture of r=0.08~arcsec radius for three different values of the background (see main text for details). The reported magnitudes include the effect of the aperture correction and the Galactic reddening correction. To correct for Galactic extinction we used A$_{\rm F225W} = 0.524$, A$_{\rm F336W} = 0.334$, A$_{\rm F555W} = 0.214$ and A$_{\rm F814W} = 0.115$. The errors reported are at the $1\sigma$ confidence level.}
\label{tab:backgroundmags}
\begin{tabular}{cccccccc}
\hline
 Filter & epoch  & min background & min background & median background & median background & max background & max background\\
 & epoch  & F$_\nu$ ($\mu$Jy) & (mag) & F$_\nu$ ($\mu$Jy) & (mag) & F$_\nu$ ($\mu$Jy) & (mag)\\
\hline 
F225W & 713 & 1.28$\pm$0.06 & 23.63 $\pm$ 0.05 & 1.18$\pm$0.06 & 23.71 $\pm$ 0.05 & 1.07$\pm$0.06 & 23.82 $\pm$ 0.06\\
F336W & 713 & 0.95$\pm$0.04 & 23.95 $\pm$ 0.05 & 0.88$\pm$0.04 & 24.02 $\pm$ 0.05 & 0.78$\pm$0.04 & 24.16 $\pm$ 0.06 \\
F555W & 713 & 0.49$\pm$0.05 & 24.68 $\pm$ 0.11 & 0.40$\pm$0.05 & 24.92 $\pm$ 0.14 & 0.30$\pm$0.05 & 25.22 $\pm$ 0.19\\
F814W & 713 & 0.57$\pm$0.12 & 24.50 $\pm$ 0.22 & 0.41$\pm$0.12 & 24.9 $\pm$ 0.3 & 0.18$\pm$0.12 & 25.8$^{+1.2}_{-0.6}$\\
\hline
F555W & 1135 & 0.42$\pm0.05$ & 24.85 $\pm$ 0.13 & 0.33$\pm$0.05 & 25.10 $\pm$ 0.17 & 0.25$\pm$0.05 & 25.41 $\pm$ 0.22\\
F814W & 1135 & 0.46$\pm$0.12 & 24.8 $\pm$ 0.4 & 0.26$\pm$0.12 & 25.4$^{+0.7}_{-0.4}$ & $<$0.34$^\dagger$ & $>$25.1$^\dagger$\\
\hline
F225W & 1474 & 0.76$\pm$0.03 & 24.19 $\pm$ 0.05 & 0.70$\pm$0.03 & 24.28 $\pm$ 0.05 & 0.65$\pm$0.3 & 24.37 $\pm$ 0.05\\
F336W & 1474 & 0.65$\pm$0.03 & 24.37 $\pm$ 0.05 & 0.61$\pm$0.03 & 24.43 $\pm$ 0.05 & 0.51$\pm$0.03 & 24.64 $\pm$ 0.07\\
F555W & 1474 & 0.40$\pm$0.05 & 24.88 $\pm$ 0.14 & 0.32$\pm$0.05 & 25.13 $\pm$ 0.17 & 0.22$\pm$0.05 & 25.53 $\pm$ 0.25\\
F814W & 1474 & 0.47$\pm$0.13 & 24.7 $\pm$ 0.3 & 0.30$\pm$0.13 & 25.2$^{+0.6}_{-0.4}$ & $<$0.43$^\dagger$ & $>$24.8$^\dagger$\\
\hline 
\end{tabular}
\newline $^\dagger$The flux density value of the background was higher than that in the aperture centred on the position of \cow, so we report the $3\sigma$ upper limit for the maximum background flux density. 
\end{table*}

Using the same 20 background regions we used for the aperture photometry on the difference images (see Figure~\ref{fig:background}), we measure the median, minimum, and maximum value for the flux density in the background aperture. There is a large spread between these three values. To investigate how the choice of background region influences the flux density measured for \cow\ we compare the results based on which of these three values is subtracted from the flux density measured in the aperture centered on the position of \cow. In Table~\ref{tab:backgroundmags} we show the resulting magnitude measurements for the different background regions. As expected, we find that using a higher background flux density yields a lower flux density for \cow. Depending on the choice of background in our work and in the work of \cite{Chen2023} both results could be consistent in all filters. We do note that in the F814W filter when using the maximum background flux density, our results  are either upper limits when the maximum background flux density was higher than the flux density in the aperture of \cow, or there are large error bars on our photometry. Clearly, the region used to determine the background flux density greatly influences the value of the magnitude of \cow. 

Next, we investigate if there are filters and epochs where the detected light originates solely from AT2018cow, or if it is possible to determine if the emission is dominated by underlying sources (for instance from diffuse emission in the galactic disk or e.g., a star forming region or cluster) 
or if it is a combination of both. Understanding the origin of the light is important because it will influence the interpretation of the power source of \cow. 

In the observations obtained through the UV filters the magnitude has decreased between epochs, suggesting that a significant fraction of the detected light is emitted by the fading transient. The SED of the light extracted at the position of \cow\, is substantially bluer than that of our compact, UV-selected, star forming regions detected throughout the host of AT2018cow. This is also in line with the notion that the majority, but not necessarily all, of the light detected in the UV arises from the transient. Subtracting a point source from the UV images at the location of \cow, leaves residuals consistent with noise (see Figure~\ref{fig:PSFresiduals}). Therefore, we conclude that the emission in the UV filters is dominated by a point source, likely the transient event \cow. In the optical filter images, comparing the observations at epoch 1 with those at epoch 3 there is evidence that the source faded in addition to light from either \cow\ (constant) and/or underlying emission from part of the host galaxy itself.

Overall, a picture emerges where light from the transient is still detected in the UV images, while in optical images we cannot determine if the detected light at epoch 3 is due to \cow\ or due to diffuse emission in the galactic disc or, more speculatively, due to a compact source at the same position of \cow. Note that in the optical images crowding is more important than in the UV images. 

The SED of the emission detected at the location of \cow\ is consistent with this picture (Figure~\ref{fig:SED}). While the F814W-F555W colour of \cow\ is consistent with that of the neighbouring sources, the F336W-F225W colour at the location of \cow\,is bluer than that of the sources in the neighbourhood. This and the fact that a single black body does not fit the SED, together with the different variability properties of the UV and optical filters, suggests that the UV and optical parts of the SED are caused by more than one emission type and/or by more than one source. This conclusion does not depend on the assumption for the nature of the light detected at 1474~days (either transient or environment light or a combination thereof). Furthermore, the emission cannot be solely from a stellar population - it is too blue - strongly implying the presence of late-time UV emission from \cow.

We also searched for BPASS single and binary stellar models, across all possible stellar ages (at Z=0.010), for models satisfying log$_{10}$(T/K)$>4.7$ and log$_{10}$(L/L$_{\odot}$)$>7.0$. These constraints are derived from fitting a black body to the late-time emission at the location of \cow\ \citep[see also][]{Sun2023}. We find no stellar models which are this blue and luminous, and therefore, a dominant contribution from an underlying massive star or binary seems ruled out by the data. 

The F555W-F814W colour at the location of \cow\ at 1474 days is $=-0.09\pm0.08$ and the absolute magnitude is $\sim -9$. Assuming that the optical bands at epoch 3 are free from light originating from the transient (as we do when taking the magnitudes measured on the difference images), we check what kind of source(s) can explain these values. They are consistent with those expected for an OB association or star-forming region \citep[e.g.,][]{Drazinos2013}, and they are broadly consistent with the F555W-F814W colours of the UV-selected, compact star-forming regions shown in Figure~\ref{fig:SF_regions}. The mean F555W-F814W colour (corrected for Galactic but not intrinsic extinction [at the specific location in the host galaxy]) of these regions is 0.02$\pm$0.05. Excluding the UV filters, fixing A$_{\rm V}=0$ and performing SED fitting as described in Section~\ref{sec:SF_regions}, we infer a best-fit population age at the location of \cow\ of 20 and 79\,Myr at 713 and 1474 days, respectively. Although we cannot determine a precise age with just two optical filters, if we assume no extinction and that the optical light is dominated by the underlying stellar population, the optical spectral slope constrains the age to $\sim$100\,Myr or less.

\begin{figure}
 \centering
 \includegraphics[width=0.99\columnwidth]{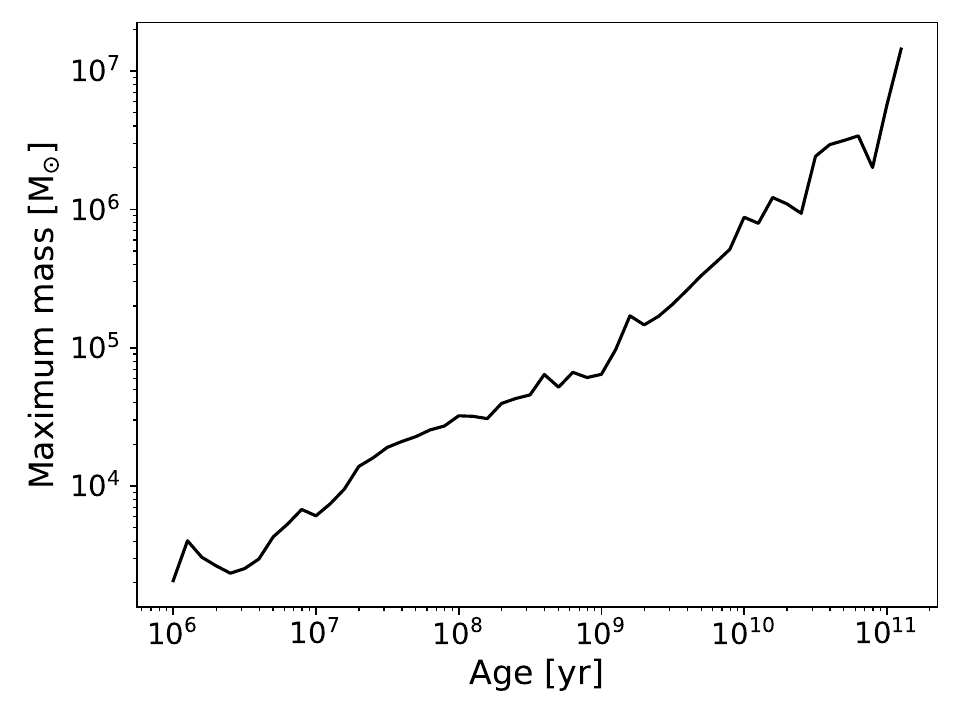}
  \caption{The maximum mass of a stellar cluster that can be underlying \cow\ as a function of population age. This is determined by the maximum luminosity of a BPASS simple stellar population that can lie at this location without the luminosity in one of the four \textit{HST} bands exceeding the observed value.}
 \label{fig:maxmass}
\end{figure}

Taking the 4-band photometry of \cow\ (latest epoch with the median background, see Table \ref{tab:backgroundmags}), and converting it to absolute magnitudes and using BPASS simple stellar populations, we calculate the maximum mass of a cluster that can be present at the location of \cow\ before the luminosity in one of the filters exceeds the magnitude plus its 1\,$\sigma$ error. We plot the upper limit on the cluster mass in Figure \ref{fig:maxmass}. This upper limit is a strong function of age - as the UV flux reduces with increasing age, the upper limit on the cluster mass increases. An old stellar population at the location of \cow\ cannot be ruled out - in particular, we note that a globular cluster can easily be hidden underneath the light of \cow\ \citep[based on typical globular cluster ages of several Gyr and masses of 10$^{3}$-10$^{6}$M$_{\odot}$,][]{1996AJ....112.1487H}. 

\subsection{Disc modelling}

It has been speculated that \cow\, is caused by a tidal disruption event (TDE; \citealp[e.g.,][]{Perley2019, Kuin2019}). Interestingly, for low mass ($M_{BH} < 10^{6.5} M_\odot$) TDEs roughly time-independent UV emission lasting for time scales of years is commonly detected (\citealt{VanVelzen2019, Mummery2020, wen2023}). Comparing the UV light curve of \cow\, (Figure~\ref{fig:lightcurve}) with that of TDEs, for example ASSASN-14li (see e.g., figure 2 in \citealt{wen2023}), we note that the UV light curve morphology is similar. Especially the late-time plateau is a distinguishing feature shared by both sources. 

To test the hypothesis if the late-time UV emission observed from AT2018cow could come from an evolving accretion flow produced by the tidal disruption of a star by a massive black hole, we follow the procedure set out in \cite{Mummery2020}, and generate evolving UV light curves by solving the time-dependent general relativistic thin disc equations. In brief, we assume that the tidal disruption of a star results in the formation of a compact ring of material with total mass roughly half that of the disrupted star. This initial ring is assumed to form at the circularisation radius (typically twice the tidal radius) of the incoming star (see also \citealt{2021ApJ...921...20H}).  Once this initial condition is specified, by solving the time-dependent relativistic disc equations, the disc density can be propagated out to large times. Once the dynamical evolution of the disc density is solved, the observed disc spectrum can be computed by ray-tracing the emergent flux of each disc annulus. This then allows us to compute late time UV luminosities for a range of black hole and stellar parameters.  

The late-time luminosity observed from the location of AT2018cow is, compared to the typical TDE population, at a relatively low level $\nu L_\nu \simeq 10^{39}$ erg/s, at $\nu \simeq 10^{15}$ Hz. This is much smaller than, for example, the luminosity of the $\sim 10^6 M_\odot$ black hole mass TDE ASASSN-14li, which had a late time ($>1$~year) UV luminosity of $\nu L_\nu \simeq 10^{42}$ erg/s. \cite{Mummery21} showed empirically from fitting the light curves of 9 TDEs that the late time UV plateau luminosity correlates approximately linearly with the black hole mass responsible for the TDE. This empirical result has strong theoretical and numerical support (Mummery \& van Velzen et al. in prep.), and suggests  that AT2018cow could well be due to a TDE involving an intermediate-mass central black hole. 

\begin{figure}
    \centering
    \includegraphics[width=0.5\textwidth]{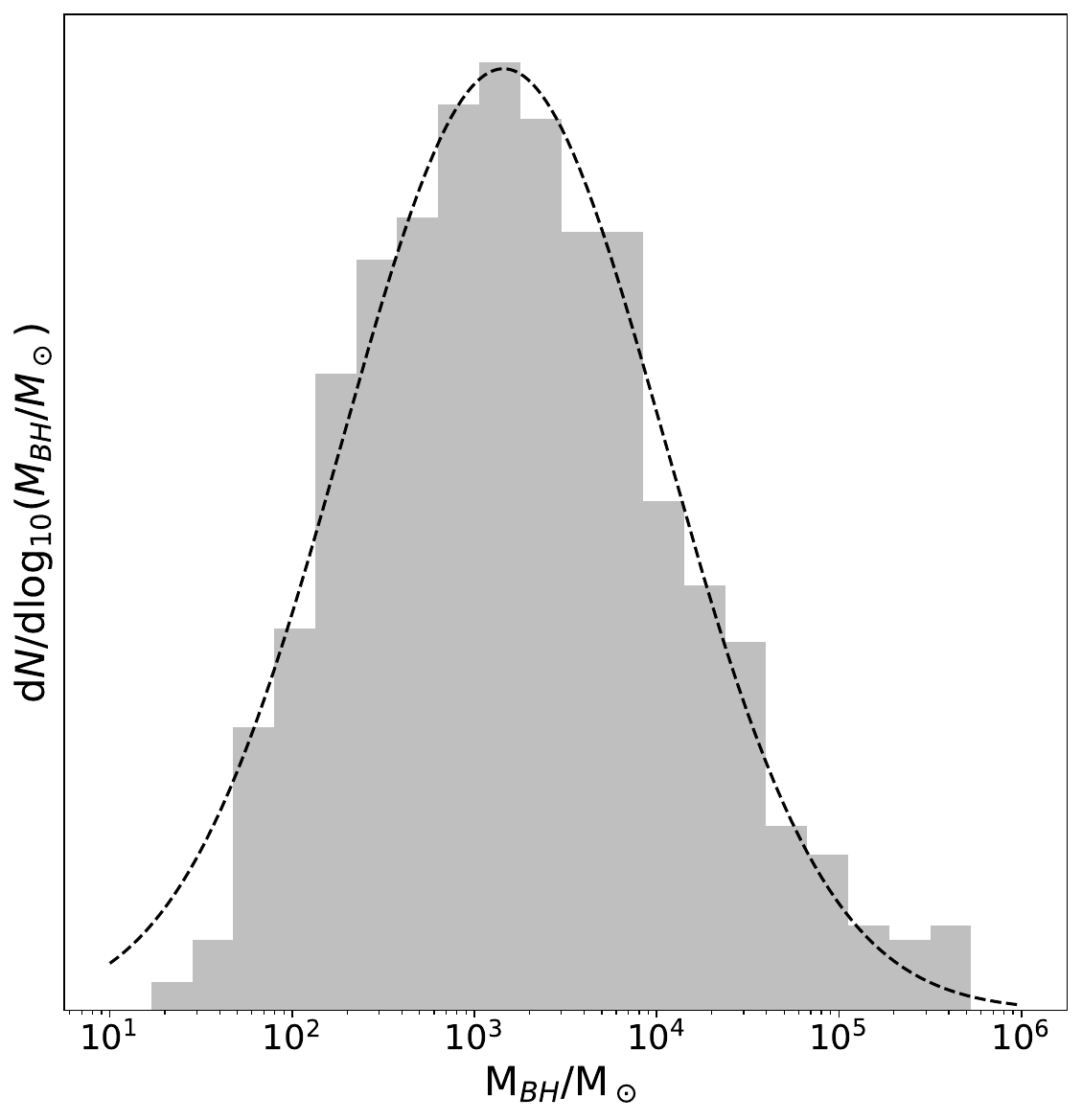}
    \caption{The black hole masses consistent with the assumption that \cow\, was caused by an tidal disruption event. The distribution of black hole masses has been derived assuming the late time UV emission is due to the accretion disc formed following the disruption (see the main text for details). The mean of the logarithm of black hole mass (M$_{\rm BH}$) is $\log$~M$_{\rm BH}$ = 3.2$\pm$0.8 (with the mass in M$_\odot$). }
    \label{fig:COWDISC1}
\end{figure}

To test this hypothesis, we numerically sample $N = 10^5$ black hole masses uniformly in the range $10^1 < M_{BH} / M_\odot < 10^7$. At each black hole mass we sample stellar masses from the Kroupa IMF (\citealt{2001MNRAS.322..231K}), solve the disc equations and ``observe'' the system at a random inclination (with $\cos(i)$ sampled uniformly). We sample uniformly the (dimensionless) black hole spin between $-1 < a < 1$. As a very conservative constraint on the central black hole mass in AT2018cow, we record all TDE disc systems which produce a UV luminosity at +713 days (the time of the first \textit{HST} observation) within a factor of 2 of $3 \times 10^{39}$ erg/s at $\nu = 10^{15}$ Hz. The black hole mass distribution of the TDE systems which satisfy this constraint are shown in Figure~\ref{fig:COWDISC1}. 

\begin{figure}
    \centering
    \includegraphics[width=0.5\textwidth]{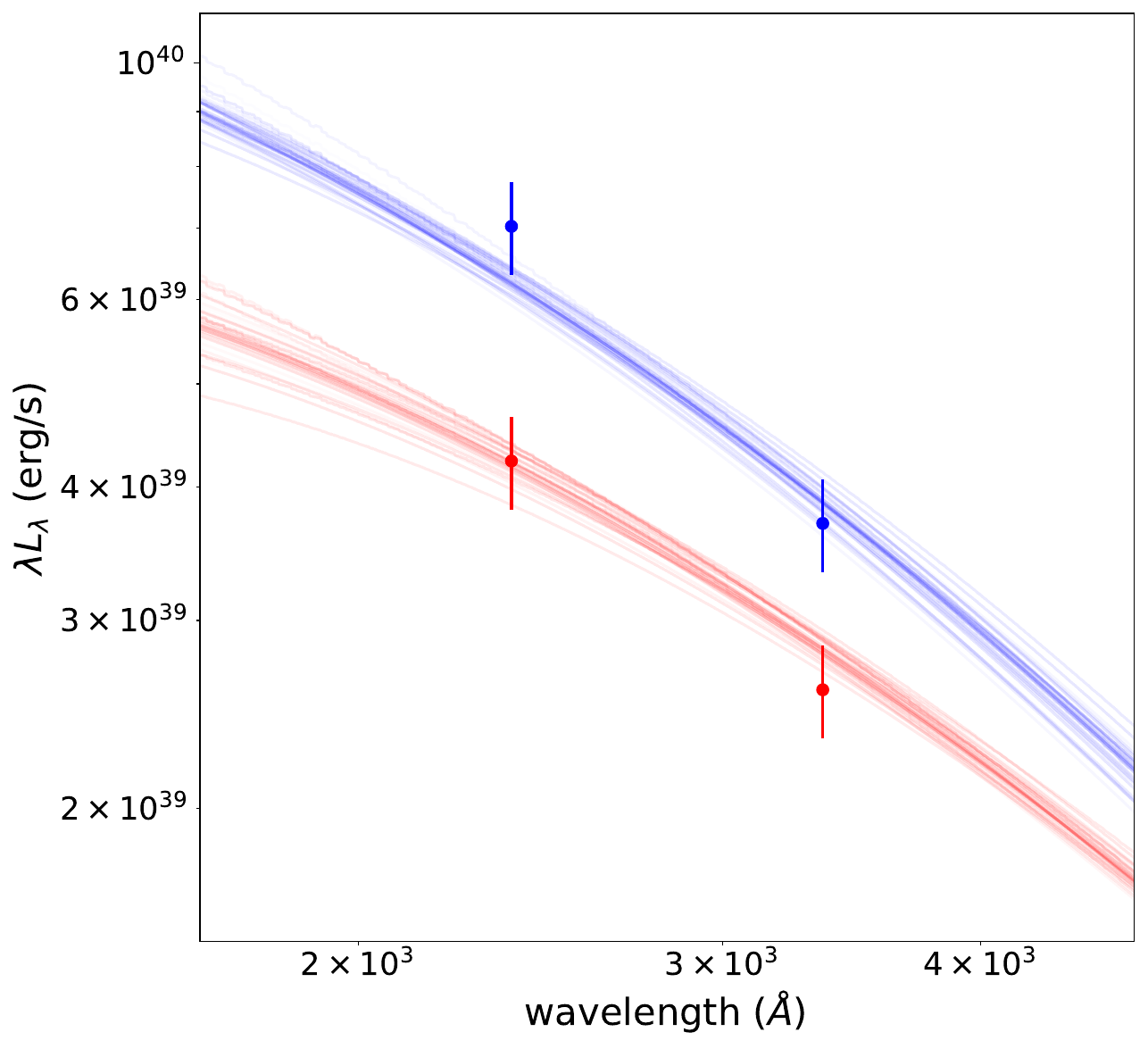}
    \caption{Late time (blue = +713 d, red = +1474 d) spectral snapshots of a sample of relativistic accretion disc models for \cow. These curves  show a sub-set of the disc models (Fig. \ref{fig:COWDISC1}) which produced particularly good fits to the data. 
    }
    \label{fig:COWDISC2}
\end{figure}

\begin{figure}
    \centering
    \includegraphics[width=0.5\textwidth]{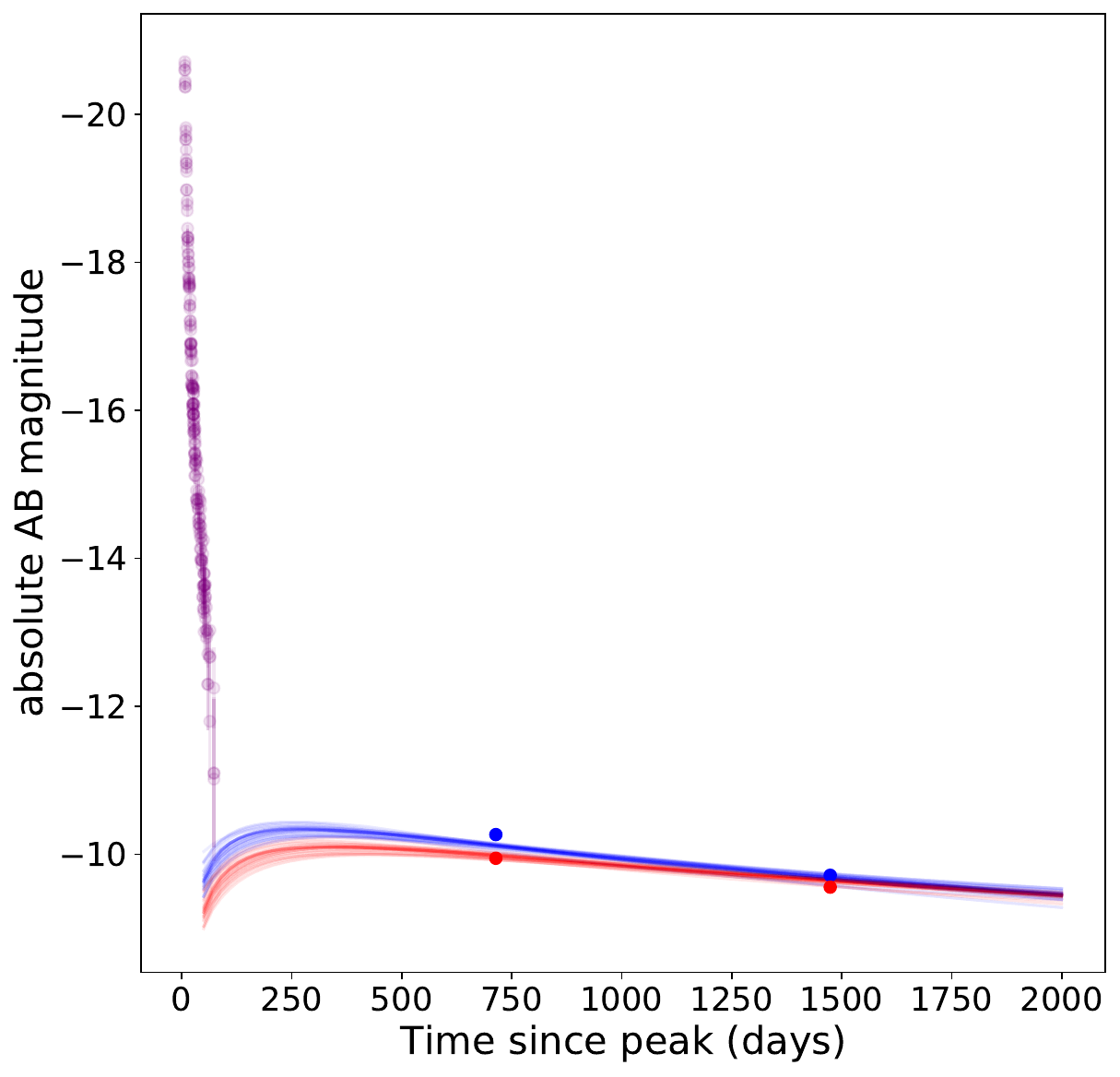}
    \caption{The light curves of the relativistic disc models which produce the spectra displayed in Figure~\ref{fig:COWDISC2}. The late time \textit{HST} data are displayed in blue (F225W) and red (F336W). Early time data in the ultra-violet bands UVW1, UVW2 and UVM2  are displayed in purple. Importantly, a disc model can reproduce the late time \cow\  UV emission, without modifying the observed early time \cow\ rapid light curve decline. There is no consensus in the TDE community about the origin of the early-time UV (and optical) emission (see, e.g., \citealt{2020SSRv..216..114R}). The error bars are at a $1\sigma$ confidence level, and may be smaller than the marker size. }
        \label{fig:COWDISC3}
\end{figure}

A more detailed analysis of the late time AT2018cow light curve and spectrum highlights that an evolving accretion flow provides a plausible explanation of the observed AT2018cow data. It is of course difficult to constrain a best fitting set of parameters from observations in two bands at two epochs, and we do not attempt to measure the precise system parameters of AT2018cow from the late time \textit{HST} data. Instead, we examine  a sub-set (200) of our solutions (Figure~\ref{fig:COWDISC1}) which produce  particularly "good fits" (as judged by their chi-squared statistic computed from both epochs). For these solutions we generate both optical-UV spectra at $t = +713$~d and $+1474$~d, and disc UV light curves from $t = 0$ to $t = +1500$~d. These disc spectra and light curves are displayed in Figures~\ref{fig:COWDISC2} and \ref{fig:COWDISC3}, respectively.  It is clear that an evolving relativistic accretion flow can reproduce the observed late-time properties of AT2018cow. 

The central black hole masses inferred from disc modelling ($M_{BH} \sim 10^{3.2\pm 0.8} M_\odot$) implies that the early-time UV/optical emission observed from AT2018cow is significantly above the Eddington luminosity L$_{\rm Edd} \simeq 10^{41} (M_{BH}/10^3M_\odot)$ erg/s.  If the early time luminosity is indeed ultimately powered by accretion (which is still uncertain, see e.g., \citealt{2020SSRv..216..114R}), then it is unlikely that the thin disc models used here would be valid at these very early times (i.e., for the first $\sim$ 100 days). However, by the much later times, which we are interested in, the bolometric luminosities of the disc solutions are typically at the level of a few $10^{40}$ erg/s (e.g., Fig. \ref{fig:COWDISC2}), suggesting Eddington ratios at the $10\%$ level, where thin disc models are valid. 

\citet{Chen2023} uses a steady state disc model to fit their T=1474~d SED and obtain an estimate for the mass for the BH. 
However, steady state disc theory predicts an optical/UV disc luminosity which scales as  $(M_{BH} \dot M)^{2/3}$. This optical/UV luminosity is thus highly degenerate between the (unknown) mass accretion rate $\dot M$, and the black hole mass $M_{BH}$ \citep[e.g.,][]{FKR2002}. However, the late time disc temperature profile in a TDE disc is  constrained, as the total initial mass, radial and temporal scales of the disc are known a priori for a given stellar disruption. This initial mass content must then propagate radially according to the standard laws of mass and angular momentum conservation. The resulting late-time optical/UV luminosity of a {\it TDE  disc} is strongly constrained. We make use of this in our disc model. 


If AT2018cow is indeed a TDE, the short time scale and the disc modelling suggests a relatively low-mass BH was responsible for the disruption. \cite{Pasham2021} find a limit of $M_{BH} < 850 M_\odot$ based on the frequency of the soft QPO. \cite{Zhang2022} find a low frequency QPO, corresponding to a BH mass of $\sim 10^4 M_\odot$ and they suggest the maximum mass found by \cite{Pasham2021} can be increased to higher mass adding a binary component to the compact object.  

A problem for the TDE hypothesis is that the BH responsible for the disruption needs to be embedded in a dense stellar environment for dynamical friction to be efficient enough to bring a star on an orbit towards its tidal radius within a Hubble time (e.g,. \citealt{2016MNRAS.455..859S}). Such a dense stellar environment where dynamical interactions occur often enough, may arise in nuclear star clusters, dense young star clusters, or globular clusters. 
There is evidence of a recent interaction between CGCG~137-068 and a companion galaxy from a ring of high column density gas as well as from a faint tidal tail \cite{Lyman2020, Roychowdhury2019}. If the host galaxy underwent a recent (minor) merger it is conceivable that an IMBH or SMBH, with its nuclear star cluster, is in the process of falling into the center of CGCG~137-068. This, means that a TDE origin of \cow\ remains a viable possibility.

However, \cite{Michalowski2019} attributes the presence of the ring of high column density gas reported by \cite{Roychowdhury2019} to internal processes instead of a galaxy merger/interaction. Their observations using H~I show no evidence for a concentration of gas near the location of \cow. They conclude that the host of \cow\, differs from the hosts of Gamma-ray bursts (GRBs)/SNs in its properties and therefore the environment of \cow\ does not provide evidence for a massive star progenitor for the event, leaving the question on the nature of \cow\,wide open.

\section{Summary and Conclusions}

Using three epochs of {\it HST} observations we investigate the late-time UV and optical emission at the location of \cow. The main results are that \cow\, remains UV-bright, even with evidence for fading in the UV filters (F225W and F336W) between the first and third epoch of {\it HST} observations. The magnitude of \cow\, in the optical filters (F555W and F814W) can differ by up to a magnitude depending on how the (diffuse galaxy) background at the location of \cow\, is determined.

From our analysis we conclude the following:
i) The observed UV emission is consistent with being dominated by a fading point source which originates most likely from \cow. ii) While part of the optical emission could be due to slowly decaying emission from the transient, there is evidence for a contribution of underlying emission, that did not fade between epochs. Some fraction of this could originate in diffuse galactic background light or an underlying point(like) source. 
iii) The late-time UV emission is reminiscent of late-time UV emission seen for TDEs. The late-time UV luminosity of \cow\ is consistent with the disruption of a (low-mass) star by an IMBH. For this scenario to be feasible \cow\, needs to reside in a dense (young/old) stellar cluster.

Our research shows that the nature of \cow\ is still uncertain. Both model scenarios involving either specific massive star evolution or a tidal disruption of a (white dwarf) star by an intermediate mass black hole have their advantages and disadvantages.

\section*{Acknowledgements}

AI thanks Luc IJspeert for helpful discussions.
This work is part of the research programme Athena with project number 184.034.002, which is financed by the Dutch Research Council (NWO).
The scientific results reported on in this article are based on data obtained under \textit{HST} Proposals 15974, 16179 and 16925 with PI A.J.~Levan, A.~Filippenko and Y.~Chen, respectively.
This work was supported by a Leverhulme Trust International Professorship grant [number LIP-202-014]. 
This work makes use of Python packages {\sc numpy} \citep{2020arXiv200610256H}, {\sc scipy} \citep{2020NatMe..17..261V}; {\sc matplotlib} \citep{2007CSE.....9...90H}, {\sc extinction} \citep{barbary_kyle_2016_804967} and {\sc drizzlepac} \citep{drizzlepack}.
This work made use of Astropy:\footnote{http://www.astropy.org} a community-developed core Python package and an ecosystem of tools and resources for astronomy \citep{astropy:2013, astropy:2018, astropy:2022}. This research made use of Photutils, an Astropy package for
detection and photometry of astronomical sources \citep{photutils}.
This research has made use of the SVO Filter Profile Service (\url{http://svo2.cab.inta-csic.es/theory/fps/}) supported from the Spanish MINECO through grant AYA2017-84089 \citep{2012ivoa.rept.1015R,2020sea..confE.182R}. 
This work has made use of v2.2.1 of the Binary Population and Spectral Synthesis (BPASS) models as described in \citet{Eldridge2017} and \citet{Stanway2018}.

\section*{Data Availability}

All data used in this paper is publicly available from the \textit{HST} data archive. A reproduction package for this paper is uploaded to Zenodo (\url{https://doi.org/10.5281/zenodo.8246571}). 



\bibliographystyle{mnras}
\bibliography{references} 




\appendix
\newpage

\section{Background regions}

\begin{figure*}
 \centering
 \hspace*{-.7cm}\includegraphics[width=0.6\textwidth]{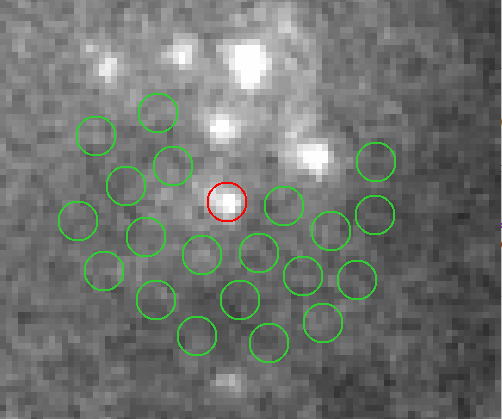}
  \caption{Greyscale F555W T=713~days image showing 20 background apertures (green) placed randomly within a distance of 30 pixels from the position of \cow, shown here in red. The background apertures are placed such to avoid bright sources in the image.}
 \label{fig:background}
\end{figure*}

\newpage

\section{Astrometry sources}

\begin{table}
\centering 
\caption{Coordinates of the 10 sources used to test the alignment between the epoch 1 F336W image and the epoch 3 F555W image in Section \ref{sec:astrometryresults}. } 
\label{apx:astrometry}
\begin{tabular}{lcc}
Source ID & RA (dd:mm:ss.ss) & Dec (dd:mm:ss.ss)\\
\hline
1 & 16:16:00.9 & +22:16:10.8 \\
2 & 16:16:00.8 & +22:16:10.4 \\
3 & 16:16:00.6 & +22:16:08.2 \\
4 & 16:16:00.9 & +22:16:08.9 \\
5 & 16:16:00.2 & +22:16:09.2 \\
6 & 16:16:00.0 & +22:16:03.7 \\
7 & 16:16:00.2 & +22:16:05.8 \\
8 & 16:16:00.2 & +22:16:06.3 \\
9 & 16:16:00.4 & +22:16:04.2 \\
10 & 16:15:59.8 & +22:15:59.2 \\
\hline
\end{tabular}
\end{table}

\newpage

\section{SED fitting results}\label{apx:bpass}
Here, we provide the results of age and extinction SED fitting to UV-selected compact star-forming regions, using BPASS synthetic spectra, as described in Section~\ref{sec:SF_regions}. In Table~\ref{tab:bpass} we provide the region ID, R.A.~and Dec., best-fit age, best-fit extinction, and reduced $\chi^{2}$, where we have 4 data points and 2 parameters yielding 2 degrees of freedom.

\begin{table*}
\centering 
\caption{Results from the SED fitting procedure described in Section \ref{sec:SF_regions}. Ages are spread logarithmically from 1\,Myr to 100\,Gyr and the extinction A$_{\rm V}$ is allowed to vary between 0 and 1 in steps of 0.1. The best-fit age and extinction for each UV-selected star-forming region, and the reduced $\chi^{2}$ of the fit, are provided. Results for the location of \cow\ (region 4) are provided in the first two rows at 713 and 1474 days respectively.} 
\label{tab:bpass}
\begin{tabular}{lccccccccc}
\hline %
Region ID & RA(hh:mm:ss.ss) & Dec(dd:mm:ss.ss) & F225W & F336W & F555W & F814W & Age/Myr & A$_{\rm V}$ & $\chi_{\nu}^{2}$ \\
\hline %
Cow (713d) & 16:16:00.2 & +22:16:04.8 & 23.73$\pm$0.05 & 24.05$\pm$0.04 & 24.92$\pm$0.04 & 24.97$\pm$0.09 & 1.0 & 0.0 & 47.0 \\ 
Cow (1474d) & 16:16:00.2 & +22:16:04.8 & 24.28$\pm$0.06 & 24.44$\pm$0.04 & 25.15$\pm$0.05 & 25.24$\pm$0.07 & 1.0 & 0.0 & 37.4 \\ 
\hline %
0 & 16:16:00.2 & +22:16:06.3 & 24.74$\pm$0.09 & 24.91$\pm$0.07 & 24.99$\pm$0.05 & 25.08$\pm$0.1 & 8.0 & 0.1 & 0.2 \\ 
1 & 16:16:00.2 & +22:16:05.8 & 25.18$\pm$0.13 & 24.81$\pm$0.07 & 24.81$\pm$0.04 & 24.46$\pm$0.06 & 10.0 & 0.3 & 3.2 \\ 
2 & 16:16:00.1 & +22:16:05.7 & 25.63$\pm$0.17 & 25.26$\pm$0.09 & 25.66$\pm$0.06 & 25.33$\pm$0.1 & 10.0 & 0.1 & 11.8 \\ 
3 & 16:16:00.2 & +22:16:05.3 & 25.24$\pm$0.17 & 24.91$\pm$0.08 & 24.42$\pm$0.03 & 23.84$\pm$0.04 & 13.0 & 0.6 & 3.1 \\ 
5 & 16:16:00.3 & +22:16:03.6 & 24.01$\pm$0.08 & 23.67$\pm$0.05 & 23.42$\pm$0.02 & 22.99$\pm$0.02 & 13.0 & 0.4 & 1.8 \\ 
6 & 16:16:00.3 & +22:16:03. & 25.44$\pm$0.16 & 25.52$\pm$0.12 & 25.76$\pm$0.07 & 25.88$\pm$0.14 & 8.0 & 0.0 & 0.9 \\ 
7 & 16:16:00.5 & +22:16:16.6 & 25.72$\pm$0.21 & 25.54$\pm$0.12 & 25.31$\pm$0.06 & 24.85$\pm$0.08 & 12.6 & 0.3 & 0.9 \\ 
8 & 16:16:00.45 & +22:16:16.5 & 25.71$\pm$0.20 & 25.87$\pm$0.15 & 25.52$\pm$0.06 & 25.07$\pm$0.08 & 12.6 & 0.3 & 1.6 \\ 
9 & 16:16:01.1 & +22:16:16.4 & 25.31$\pm$0.15 & 24.91$\pm$0.08 & 25.25$\pm$0.05 & 24.98$\pm$0.08 & 10.0 & 0.1 & 10.8 \\ 
10 & 16:16:01.0 & +22:16:16.3 & 25.11$\pm$0.12 & 24.88$\pm$0.07 & 25.24$\pm$0.05 & 25.68$\pm$0.12 & 1.3 & 0.4 & 2.8 \\ 
11 & 16:16:00.4 & +22:16:15.1 & 26.09$\pm$0.26 & 25.60$\pm$0.11 & 25.99$\pm$0.07 & 25.55$\pm$0.11 & 10.0 & 0.1 & 7.7 \\ 
12 & 16:16:00.2 & +22:16:13.3 & 25.91$\pm$0.23 & 25.76$\pm$0.13 & 25.64$\pm$0.07 & 25.02$\pm$0.09 & 12.6 & 0.3 & 3.8 \\ 
13 & 16:16:00.9 & +22:16:12.2 & 26.01$\pm$0.23 & 25.50$\pm$0.11 & 26.08$\pm$0.09 & 26.69$\pm$0.30 & 1.3 & 0.4 & 3.6 \\ 
14 & 16:16:00.1 & +22:16:11.2 & 25.50$\pm$0.16 & 24.93$\pm$0.08 & 25.30$\pm$0.05 & 25.30$\pm$0.09 & 7.9 & 0.2 & 10.6 \\ 
15 & 16:16:01.0 & +22:16:10.7 & 25.49$\pm$0.18 & 24.92$\pm$0.08 & 24.64$\pm$0.04 & 24.51$\pm$0.06 & 8.0 & 0.7 & 0.5 \\ 
16 & 16:16:00.8 & +22:16:10.4 & 24.77$\pm$0.10 & 24.74$\pm$0.07 & 24.89$\pm$0.04 & 24.87$\pm$0.07 & 7.9 & 0.2 & 3.0 \\ 
17 & 16:16:00.3 & +22:16:09.4 & 25.50$\pm$0.18 & 24.93$\pm$0.09 & 25.46$\pm$0.06 & 26.45$\pm$0.25 & 1.3 & 0.4 & 6.4 \\ 
18 & 16:16:01.0 & +22:16:08.8 & 25.34$\pm$0.16 & 24.86$\pm$0.08 & 25.03$\pm$0.05 & 24.79$\pm$0.08 & 10.0 & 0.2 & 6.3 \\ 
19 & 16:16:00.5 & +22:16:08.1 & 25.78$\pm$0.20 & 25.55$\pm$0.11 & 25.55$\pm$0.08 & 25.80$\pm$0.21 & 6.3 & 0.4 & 0.2 \\ 
20 & 16:16:01.2 & +22:16:03.2 & 24.22$\pm$0.08 & 24.19$\pm$0.06 & 24.35$\pm$0.03 & 24.27$\pm$0.05 & 10.0 & 0.0 & 6.6 \\ 
21 & 16:16:01.2 & +22:16:03.0 & 25.74$\pm$0.20 & 25.54$\pm$0.12 & 25.46$\pm$0.06 & 25.22$\pm$0.10 & 10.0 & 0.2 & 0.6 \\ 
22 & 16:16:00.7 & +22:16:00.9 & 26.18$\pm$0.29 & 25.58$\pm$0.12 & 25.69$\pm$0.06 & 25.14$\pm$0.08 & 12.6 & 0.3 & 5.5 \\ 
23 & 16:16:00.5 & +22:16:00.4 & 25.48$\pm$0.17 & 25.40$\pm$0.11 & 25.34$\pm$0.05 & 25.34$\pm$0.09 & 7.9 & 0.3 & 0.2 \\ 
\hline 
\end{tabular}
\end{table*}


\bsp	
\label{lastpage}
\end{document}